\documentclass{aa}  

\usepackage{graphicx}
\usepackage{amssymb,amsmath}
\usepackage{hyperref}			
\usepackage{lmodern}
\usepackage{mathtools}
\usepackage{comment}
\usepackage{color}
\usepackage[normalem]{ulem}
\usepackage[varg]{txfonts}

\makeatletter
\renewcommand*\aa@pageof{, page \thepage{} of \pageref*{LastPage}}
\makeatother

\begin{document}
\title{\textcolor{black}{The low surface thermal inertia of the rapidly rotating near-Earth asteroid 2016 GE1}}

\titlerunning{Thermal characteristics of asteroid 2016 GE1}

   \author{Marco Fenucci\inst{1,2,3}
             \and
          Bojan Novakovi\'c\inst{2}
          \and
          Du\v{s}an Mar\v{c}eta\inst{2}
          }
   \authorrunning{Fenucci et. al.}
    \institute{ESA ESRIN / PDO / NEO Coordination Centre, Largo Galileo Galilei, 1, 00044 Frascati, RM, Italy\\
    \email{marco.fenucci@ext.esa.int}
    \and
    Department of Astronomy, Faculty of Mathematics, University of Belgrade,
             Studentski trg 16, 11000 Belgrade, Serbia
                 \and
    Elecnor Deimos, Via Giuseppe Verdi, 6, 28060 San Pietro Mosezzo, NO, Italy
             }
   \date{Received --- / Accepted ---}


  \abstract
     {Asteroids smaller than about 100 meters are observed to rotate very fast, with periods often much shorter than the critical spin limit of 2.2 h. Some of these super-fast rotators can also achieve a very large semi-major axis drift induced by the Yarkovsky
  effect, that in turn, is determined by internal and surface physical properties.}
     {We consider here a small super-fast rotating near-Earth asteroid, designated as 2016 GE1.
     This object rotates in just about 34 seconds, and a large Yarkovsky effect has been
     determined from astrometry. By using those results, we aim to constrain the thermal inertia of the surface of this
     extreme object.}
     {We used a recently developed statistical method to determine the thermal
     properties of near-Earth asteroids. The method is based on the comparison between
     the observed and the modelled Yarkovsky effect, and the thermal conductivity (inertia) is
     determined by a Monte Carlo approach. Parameters of the Yarkovsky effect model are either
     fixed if their uncertainty is negligible, modelled with a
     Gaussian distribution of the errors if they are measured, or deduced from general
     properties of the population of near-Earth asteroids when they are unknown.}
  {Using a well-established orbit determination procedure, we determined the Yarkovsky effect on 2016 GE1, and verified a significant semi-major axis drift rate. Using a statistical method, we showed that this semi-major axis drift rate could be explained only by low thermal inertia values below 100 J m$^{-2}$ K$^{-1}$ s$^{-1/2}$. We benchmarked our statistical method using the well-characterised asteroid Bennu and found that the sole knowledge of semi-major axis drift rate and rotation period is generally insufficient to determine the thermal inertia. However, when the statistical method is applied to super-fast rotators, we found that the measured Yarkovsky effect can be achieved only for very low values of thermal inertia: namely, 90\% of the probability density function of the model outcomes is contained at values smaller than 100 J m$^{-2}$ K$^{-1}$ s$^{-1/2}$.} 
     {We propose two possible interpretations for the extremely low thermal inertia of 2016~GE1: a high porosity or a cracked surface, or a thin layer of fine regolith on the surface. Though this seems somewhat unexpected in either case, it opens up the possibility of a subclass of low inertia, super-fast rotating asteroids.}



   \maketitle

\section{Introduction}
\label{sec:intro}

Understanding the physical properties of asteroids is required for modelling many processes, including space weathering, formation of planetesimals, the entry of bolides into planetary atmospheres, granular mechanics, impact cratering and the thermal evolution of their parent bodies, activity drivers, and many others \citep[e.g.][]{2018ChEG...78..269F}. They are also a key to properly modelling the long-term evolution of collisional asteroid families \citep{2022CeMDA.134...34N}. Besides, insight into the physical properties of asteroids is essential for the design of robotic, lander, and sample return spacecraft missions to small bodies \citep{2021MNRAS.503.3460M}.

Despite their great importance, little is known about the surface and internal properties of asteroids because most of them are difficult to constrain from remote observations. For instance, surface properties, such as cohesion and porosity, can be deduced by estimating thermal inertia. These, however, require infrared observations \citep{2020A&A...638A..84A}, which are generally difficult to perform for small asteroids. Consequently, a reliable estimation of the thermal inertia is available only for a limited number of objects
\citep[see e.g.][]{delbo-etal_2007, delbo-etal_2015,harris-drube_2016, 2019A&A...625A.139M}. Though the situation starts changing \citep[][]{2021PSJ.....2..161M,2022PSJ.....3...56H}, the new data on asteroid thermal properties and alternative methods for their determination are still of great importance.

Recently, \citet{2021A&A...647A..61F} proposed a statistical method to estimate the surface thermal conductivity of near-Earth asteroids (NEAs). The procedure is based on the comparison between the model predicted and the measured values of the Yarkovsky effect, and, as such, it relies mainly on ground-based observations. We recall here that a generally similar idea was proposed by \citet[][see also \citet{2011MNRAS.415.2042R}]{2014A&A...568A..43R}. These authors estimated the thermal properties of asteroids with a given Yarkovky drift using the Thermophysical Model (TPM). However, though the approaches share some conceptual similarities, the method based on TPM requires data such as shape models and thermal-infrared observations. Therefore, it can provide accurate estimates of the thermal properties of individual asteroids, but for a limited number of objects for which necessary information are available. Our Monte Carlo (MC) model presented here, on the other hand, generally has limited accuracy for individual objects, if only population-based parameters are used.

In addition to that, the MC model may work well also in certain individual cases, such as extremely fast rotators. The key point here is that the magnitude of the Yarkovsky effect depends on a temperature gradient across the surface. For fast-rotating objects, such a gradient could be present only in case of low surface thermal inertia, which provides an additional constraint of the model, allowing the reliable estimation of the thermal inertia.  This opportunity was already used by \citet{2021A&A...647A..61F}, to estimate the thermal properties of the small super-fast rotating NEA (499998) 2011 PT. Here we followed the same logic and estimated the thermal inertia of another super-fast rotating object, namely 2016 GE1. 

The low thermal inertia of a small super-fast rotating asteroid (499998) 2011 PT found by \citet{2021A&A...647A..61F} was generally unexpected. Such findings may point to either a ruble-pile internal structure or the presence of a dust layer at the surface.
Though generally possible, both scenarios are relatively unexpected for super-fast rotating bodies. 
Asteroids with a rotation period shorter than 2.2 hours are known, but they are typically small, and therefore, they are thought to be rocky, monolithic asteroids. This is because strength-less objects, such as rubble-piles, should start disintegrating once their rotation period approaches the rotational disruption limit of 2.2 hours \citep{pravec-harris_2000}. Still, this theory has known exceptions, suggesting that even ruble-pile asteroids are not completely strengthless. For instance, \citet{2021Icar..36214433Z} studied the asteroid (65803) Didymos\footnote{The asteroid (65803) Didymos is the target of the first asteroid deflection test (NASA's Double Asteroid Redirection Test, DART) and the first binary asteroid system that will be characterized by a rendezvous mission (ESA's Hera).} and showed that it should have a bulk cohesion on the order of at least 10 Pa in order to maintain its structural stability. 
Regarding the presence of a dust layer on the surface of fast-spinning asteroids, theoretical works suggest that even such objects could maintain small dust particles and gravel on their surface under relatively weak cohesion \citep[][]{sanchez-scheeres_2020}. However, we do not have any direct evidence for it yet. In this respect, the extended Hayabusa~2 mission is planned to rendezvous with asteroid 1998~KY26, a small super-fast rotating asteroid, in 2031 \citep{2021AdSpR..68.1533H}, and it will provide a better understanding of these intriguing objects.

Building upon the previous works and results, here we studied in detail asteroid 2016~GE1, another super-fast rotating NEA that shares some similarities with asteroid 2011~PT studied in \citep{2021A&A...647A..61F}. Both objects are small and rotate extremely fast: 2011~PT in about 10 minutes and 2016~GE1 in only 34 seconds. Interestingly, we constrained the thermal inertia of 2016~GE1 to extremely low values with high probability, similar to the case of 2011~PT. As mentioned above, these findings point out a subpopulation of super-fast rotating asteroids that are either dust-covered or of high micro-porosity. However, we have also discussed possible alternative explanations.

The paper is organized as follows. In Sec.~\ref{s:OD} we report the Yarkovsky effect measurement obtained by orbit determination. In Sec.~\ref{s:methods} we described the method used for the estimation of the thermal properties, while in Sec.~\ref{sec:testing}, basic testing and verification of the model is presented. The results of the method applied to 2016~GE1 are reported in Sec.~\ref{s:results}. In Sec.~\ref{s:dis} we discuss objects similar to 2016~GE1 and the implications of our results. Finally, we summarize our conclusions in Sec.~\ref{s:conclusions}.

\section{Preliminary consideration}

\subsection{Orbit determination and Yarkovsky effect detection}
\label{s:OD}
\begin{table*}
   \centering
   \caption{The nominal osculating orbital elements of 2016~GE1 and their corresponding uncertainties at epoch 59800 MJD. The second and third columns show the orbit of the JPL SBDB, while the fourth and fifth columns report the solution computed with \texttt{OrbFit}.}
   \label{tab:orb_param}
   \begin{tabular}{cccccc}
      \hline
      \hline
      Parameter   & Value (JPL)  & $1\sigma$ uncertainty (JPL) & Value (\texttt{OrbFit)} & $1\sigma$ uncertainty (\texttt{OrbFit)} & Units\\
      \hline
      $a$      &  \hphantom{00}2.06282123907  & $1.2333 \times 10^{-7}$ &  \hphantom{00}2.06282138428  & $7.6035 \times 10^{-8}$     &  au\\
      $e$      &  \hphantom{00}0.52018004318  & $9.8657 \times 10^{-8}$ &  \hphantom{00}0.52018002178  & $6.2179 \times 10^{-8}$     & /  \\
      $i$      &  \hphantom{0}10.72841738208  & $1.3541 \times 10^{-5}$ &  \hphantom{0}10.72841426920  & $1.3243 \times 10^{-5}$     & deg\\
      $\Omega$ &  \hphantom{0}15.61505016189  & $2.9292 \times 10^{-6}$ &  \hphantom{0}15.61505197650  & $1.6644 \times 10^{-6}$     & deg\\
      $\omega$ &             197.28049788594  & $1.0439 \times 10^{-4}$ &             197.28045820348  & $6.6534 \times 10^{-5}$     & deg\\
      $M$   &  \hphantom{0}41.65276790391  & $4.2564 \times 10^{-5}$ &  \hphantom{0}41.65274693145  & $2.6363 \times 10^{-5}$     & deg\\
      $A_2$    & $-1.4388655 \times 10^{-12}$ & $4.3780 \times 10^{-13}$& $-9.6496757 \times 10^{-13}$ & $2.8004 \times 10^{-13}$    & au d$^{-2}$\\
      $(\text{d}a/\text{d}t)_{\text{m}}$ &  $-0.05832$ & $0.01775$ &  $-0.03912$ & $0.01136$ & au My$^{-1}$ \\
      \hline
   \end{tabular}
\end{table*}
The orbital parameters of 2016~GE1 provided by the JPL Small-Body Database\footnote{\url{https://ssd.jpl.nasa.gov/}} (SBDB) are reported in Table~\ref{tab:orb_param}, together with their uncertainties. 
We also independently performed the orbit determination by using the \texttt{OrbFit} free software\footnote{\url{http://adams.dm.unipi.it/orbfit/}}, version 5.0.8. 
The Minor Planet Center\footnote{\url{https://www.minorplanetcenter.net/}} reports a total number of 127 observations for 2016~GE1, obtained during the nights of 2016 April 2$^{\text{nd}}$ and 2019 April 5$^{\text{th}}$, coinciding to the close approaches with the Earth happened at a distance of 0.00356 au and 0.00743 au, respectively. All the available observations were used for the computation of the orbit. %
The dynamical model used for the orbital fit is similar to the one described in \citet{2018A&A...617A..61D}, and includes the gravitational forces of the Sun, the eight planets, the Moon, the 16 most massive main-belt asteroids, and Pluto. The masses and the positions of these bodies are all computed by using the JPL ephemerides DE431 \citep{2014IPNPR.196C...1F}. In addition, we added the relativistic effects of the Sun, the planets, and the Moon expressed as a first-order post-Newtonian expansion. The Yarkovsky effect is modeled as in \citet{farnocchia-etal_2013}, as an acceleration along the transverse direction of motion $\hat{\mathbf{t}}$ of the form
\begin{equation}
    \mathbf{a}_t = \frac{A_2}{r^2}\hat{\mathbf{t}},
\end{equation}
where $r$ is the distance from the Sun. The parameter $A_2$ is determined together with the orbital elements by fitting the model to the observations through a least-square procedure \citep[see, e.g.][]{milani-gronchi_2009}. The orbit determination algorithm is also endowed with an automatic outlier rejection procedure, described in \citet{2003Icar..166..248C}. 

The orbital elements and the value of $A_2$ obtained with our orbital fit are reported in Table~\ref{tab:orb_param}. Of the total 127 observations provided by the MPC, only 1 was rejected as an outlier. The weighted Root Mean Square (RMS) of the astrometric residuals resulted to be 0.522 arcsec, which is only slightly smaller than the RMS of 0.567 arcsec obtained by fitting the orbit of 2016~GE1 without the Yarkovsky effect. 
The orbital parameters that we determined are in good agreement with those provided by the JPL SBDB, in the sense that they are the same within $2\sigma$. Note also that \texttt{OrbFit} provides slightly smaller uncertainties. The values of $A_2$ also agree within $2\sigma$ uncertainty, with \texttt{OrbFit} giving the smallest nominal semi-major axis drift. The value of the signal-to-noise ratio is 3.3 for the JPL solution, and 3.4 for the \texttt{OrbFit} solution, suggesting that the detection of the Yarkovsky effect is positive. The semi-major axis drift associated to the value of $A_2$ obtained from astrometry is also reported in Table~\ref{tab:orb_param}. 

\subsection{Preliminary constraints on the 2016 GE1's thermal inertia}

In the previous sub-section, we verified the positive detection of the Yarkovsky effect and the corresponding induced drift in the semi-major axis $\text{d}a/\text{d}t$, which was found to be of considerable magnitude. This situation is somewhat unusual for a super-fast rotator such as the 2016 GE1, as the Yarkovsky effect mechanism requires a temperature gradient across the surface in order to be effective. Therefore, only specific surface thermal properties may be able to produce the measured semi-major axis drift.  

For this reason, we performed rough preliminary constraints on surface thermal inertia. To this purpose, we use the analytical
implementation of the Yarkovsky effect in our model \citep[see][]{2021A&A...647A..61F}. Instead of using the entire distribution of the input parameters, we defined some extreme values and provided each parameter as a single constant value. This allows for constraining the whole range of possible thermal inertia.   

\begin{table*}[]
    \caption{The maximum plausible range for the parameters of asteroid 2016 GE1 relevant for the magnitude of the Yarkovsky effect.}
    \centering
    \begin{tabular}{lccccc}
         \hline
    Parameter & Nominal value & Min. value & Max. value & Units  & Note \\
         \hline
    Absolute magnitude, $H$ & 26.7  & 26.0  & 27.4  &  mag  & Assumed uncertainty of 0.5 mag \\
    Geometric albedo, $p_V$ &  - & 0.04  & 0.44  & & Assumed range \\
    Bond albedo, $A$ &  - & 0.016  & 0.173 &  & Derived from $p_V$ assuming $G=0.15$\\
    Density, $\rho$   &  - & 1000  & 3500  & kg m$^{-3}$ & Assumed range \\
    Rotation period, $P$ & 34  & 17  & 340  & sec & Assumed range \\
    Diameter, $D$ &  - &  6 &  41 & m & Derived from the range of $H$ and $p_v$\\
    Yarkovsky drift, $da/dt$  &  -  & -0.027  & -0.076  & au My$^{-1}$ & Based on the values from Table~\ref{tab:orb_param}\\
         \hline
    \end{tabular}
    \label{tab:GE1_extreme_range}
\end{table*}

In order to define the range of possible values of the surface thermal inertia of 2016 GE1, we tried different combinations of the
minimum and maximum values of the parameters given in Table~\ref{tab:GE1_extreme_range}.  By doing so, we considered that some parameters are correlated with changes in thermal inertia, while others are anti-correlated. For instance, larger values of the Yarkovsky drift are compatible with lower values of the thermal inertia and vice-versa. As the drift is inversely proportional to the mass of the object, increasing the size or density of the asteroid, while keeping the other parameters fixed, results in lower thermal inertia.

By estimating the thermal inertia from analytical Yarkovsky formulation, we found that the thermal inertia solution is not possible for many of the combinations of the extreme values of the parameters. Nevertheless, we found a number of successful estimations.
We highlight that in all of these solutions, we obtained $\Gamma < 50$ m$^{-2}$ K$^{-1}$ s$^{-1/2}$, which is a strong indication that the surface thermal inertia of 2016 GE1 is very low. Building on this interesting indication, we used the more complex statistical model to better constrain the thermal inertia of GE1. This model also relies on the semi-analytic implementation of the Yarkovsky effect that takes into account the orbital eccentricity (see Section~\ref{s:methods}), which is very important for the considered object (see Table~\ref{tab:orb_param}).      

\section{Monte Carlo model and input parameters}
\label{s:methods}
We use the Monte Carlo method developed by \citet{2021A&A...647A..61F} to estimate the thermal properties of 2016~GE1. The method is based on the comparison between the Yarkovsky drift measured from astrometry 
\citep[see e.g.][]{farnocchia-etal_2013, 2018A&A...617A..61D, greenberg-etal_2020},
and the model-predicted value \citep[see e.g.][]{1999A&A...344..362V, bottke-etal_2006, vokrouhlicky-etal_2017}. 

In practice, the MC model by \citet{2021A&A...647A..61F} searches for input parameters so that a theoretically predicted value of the Yarkovsky effects best matches a measured value. In this respect, we recall that the Yarkovsky effect depends on several orbital and physical parameters: the orbital semi-major axis $a$, the orbital eccentricity $e$, the diameter $D$, the density $\rho$, the thermal conductivity $K$, the heat capacity $C$, the obliquity $\gamma$, the rotation period $P$, the absorption coefficient $\alpha$, and the emissivity $\varepsilon$. Suppose all but one parameter are fed to the model as the inputs. In that case, the remaining parameter could be determined, provided that a measurement $(\text{d}a/\text{d}t)_{\text{m}}$ of the Yarkovsky effect is available.

Among all the parameters, the thermal conductivity $K$ is the most uncertain one, because it strongly depends on the type of materials present at the surface of the asteroid, and it can vary by several orders of magnitude \citep{delbo-etal_2015}. Therefore, it is the one to be determined by the model. The model vs observed Yarkovsky drift equation 
\begin{equation}
   \left(\frac{\text{d}a}{\text{d}t}\right) (a, e, D, \rho, K, C, \gamma, P, \alpha,
   \varepsilon) = \bigg(\frac{\text{d}a}{\text{d}t}\bigg)_\text{m},
    \label{eq:yarkoInvertFormula}
\end{equation}
is solved for $K$ on a set of parameters randomly sampled from the input distributions, and a probability density function (PDF) is reconstructed from the output sample of $K$. 
\subsection{Semi-analytical Yarkovsky model}
\label{s:yarko_model}
In \citet{2021A&A...647A..61F}, the left-hand side of Eq.~\eqref{eq:yarkoInvertFormula} was computed by using the analytical model by \citet{1999A&A...344..362V}, that assumes a spherical shape of the asteroid, a circular orbit, and a linearization of the surface boundary condition. 
Despite the circular model being appropriate for (499998) 2011~PT, this may not be the case for many objects for which the Yarkovsky effect has been determined through astrometry, because NEAs generally reside at moderately to high eccentricity orbits.  

For this reason, here we implemented a Yarkovsky model that takes into account the effect of the eccentricity in the orbit of the asteroid. 
The instantaneous osculating semi-major axis drift caused by the Yarkovsky effect is given by
\begin{equation}
    \bigg(\frac{\text{d}a}{\text{d}t}\bigg)_{\text{i}} = \frac{2}{n^2 a} \mathbf{f}_{\text{Y}} \cdot \mathbf{v},
    \label{eq:dadt_i}
\end{equation}
where $a$ is the semi-major axis of the asteroid orbit, $n$ is the mean motion, $\mathbf{v}$ is the heliocentric orbital velocity, and $\mathbf{f}_{\text{Y}}$ is the instantaneous value of the Yarkovsky acceleration.
The term $\mathbf{f}_{\text{Y}} $ is computed by an analytical model described in \citet{vokrouhlicky-etal_2017}, that assumes a spherical shape of the asteroid and a linearization of the surface boundary condition, and it is given by 
\begin{equation}
     \mathbf{f}_{\text{Y}} = \mathbf{f}_{\text{Y, d}} + \mathbf{f}_{\text{Y, s}},
     \label{eq:yarko_add}
\end{equation}
where $\mathbf{f}_{\text{Y, d}}, \mathbf{f}_{\text{Y, s}}$ are the diurnal and the seasonal component, respectively.
The diurnal component is expressed as
\begin{equation}
    \mathbf{f}_{\text{Y, d}} = \kappa [(\mathbf{n} \cdot \mathbf{s})\mathbf{s} + \gamma_1 (\mathbf{n} \times \mathbf{s}) + \gamma_2 \mathbf{s} \times (\mathbf{n} \times \mathbf{s})].
    \label{eq:f_Yd}
\end{equation}
In Eq.~\eqref{eq:f_Yd}, $\mathbf{n} = \mathbf{r}/r$ is the heliocentric unit position vector, and $\mathbf{s}$ is the unit vector of the asteroid spin axis. In addition,
\begin{equation}
    \kappa = \frac{4 \alpha}{9} \frac{SF}{m c},
\end{equation}
where $S = 4\pi R^2$ is the cross section of the asteroid, $R$ is the radius, $F$ is the solar radiation flux at a heliocentric distance $r$, $m$ is the asteroid mass, and $c$ is the speed of light. The coefficients $\gamma_1, \gamma_2$ are expressed as
\begin{equation}
    \begin{split}
        \gamma_1 & = -\frac{k_1 \Theta}{1 + 2 k_2 \Theta + k_3 \Theta^2},\\
        \gamma_2 & = -\frac{1 + k_2 \Theta}{1 + 2k_2 \Theta + k_3 \Theta^2}, 
    \end{split}
    \label{eq:gamma1gamma2}
\end{equation}
where $\Theta = \sqrt{\rho K C \omega_{\text{rot}}}/(\varepsilon \sigma T_\star^3)$ is the thermal parameter, $\omega_{\text{rot}}$ is the rotation frequency, $\sigma$ is the Stefan-Boltzmann constant, and $T_\star$ is the sub-solar temperature, which is given by $4\varepsilon \sigma T_\star^4 = \alpha F$. The coefficients $k_1, k_2,$ and $k_3$ are positive analytic functions of the rescaled radius $R'_{\text{d}} = R/l_{\text{d}}$, where $l_{\text{d}} = \sqrt{K/(\rho C \omega_{\text{rot}})}$ is the penetration depth of the diurnal thermal waves. The analytic expressions of the $k_i, i=1,2,3$ coefficients can be found in \citet{vokrouhlicky_1998, 1999A&A...344..362V}.

The seasonal component is given by
\begin{equation}
    \mathbf{f}_{\text{Y, s}} = \kappa [ \bar{\gamma}_1 (\mathbf{n} \cdot \mathbf{s}) + \bar{\gamma}_2 (\mathbf{N}\times \mathbf{n})\cdot \mathbf{s} ]\mathbf{s},  
\end{equation}
where $\mathbf{N}$ is the unit vector normal to the orbital plane, and $\bar{\gamma}_1, \bar{\gamma}_2$ have the same expressions as Eq.~\eqref{eq:gamma1gamma2}, but evaluated with thermal parameter $\bar{\Theta} = \sqrt{\rho K C n}/(\varepsilon \sigma T_\star^3)$, and with rescaled radius $R'_{\text{s}} = R/l_{\text{s}}$ where $l_{\text{s}} = \sqrt{K/(\rho C n)}$ is the penetration depth of the seasonal thermal waves.

The average Yarkovsky drift $\text{d}a/\text{d}t$ is then obtained by averaging the instantaneous Yarkovsky drift of Eq.~\eqref{eq:dadt_i} over an orbital period, i.e.
\begin{equation}
    \frac{\text{d}a}{\text{d}t} = \frac{1}{2\pi}\int_0^{2\pi} \bigg(\frac{\text{d}a}{\text{d}t}\bigg)_{\text{i}} \, \text{d}\ell,
    \label{eq:dadt_semianalytical}
\end{equation}
where $\ell$ is the mean anomaly. The integral at the right-hand side of Eq.~\eqref{eq:dadt_semianalytical} is numerically computed with the trapezoid rule \citep[see e.g.][]{bulirsch:02}. To this purpose, we used a fixed step in the eccentric anomaly $u$, which is then translated into a step in $\ell$. This is done to secure a proper sampling of the orbit around the perihelion.

\subsection{Defining the input parameters}
\label{ss:2016GE1_param}

The next step towards modelling the Yarkovsky effect is to generate the most likely probability distribution of the input parameters on which the effect depends. As the availability of these parameters and their uncertainties could be very different, we divide the input parameters into three categories. 

The first group includes the parameters known with high accuracy. In such cases, instead of providing the probability distribution of a parameter to the model, we used only a single value, which is the nominal value of the parameter. The second group contains parameters that are determined for the individual asteroids, but their uncertainties are not negligible. Such parameters are modelled assuming a Gaussian distribution with a standard deviation corresponding to the estimated errors of the parameters. Finally, the third group includes the parameters which are, in most cases, not available for individual objects. In these cases, their probability distribution is derived from a population-based distribution.

\subsubsection{Fixed parameters}
Changes within $3\sigma$ in the semi-major axis $a$ and eccentricity $e$ produce negligible fluctuations in the left-hand side of Eq.~\eqref{eq:yarkoInvertFormula}, and therefore they are kept at their nominal values. On the other hand, the heat capacity $C$, the emissivity $\varepsilon$, and the absorption coefficient $\alpha$ are all unknown. However, the plausible range for each of these parameters is narrow compared to the uncertainty of other relevant quantities. Therefore, instead of providing a full distribution of these parameters, we fixed their values in each simulation and provided results with a few different values.
Typical heat capacity values assumed for asteroids are in the range 600$-$1200 J kg$^{-1}$ K$^{-1}$ \citep{farinella-etal_1998, delbo-etal_2015, 2021JGRE..12607003P}. Therefore we performed simulations for a few values from this interval. 

For the emissivity $\varepsilon$, we adopted $0.984$ as a nominal value corresponding to the mean value of measurements performed on meteorites \citep{ostrowsky-bryson_2019}. Additionally, we have tested how much the results change if a value of $\varepsilon$ = $0.9$ is assumed.

The absorption coefficient is defined as $\alpha=1-A$, where $A$ is the Bond albedo of the object. As we found that in the case of 2016 GE1, it does not affect the main conclusions (see Section~\ref{ss:rubustness}), the absorption coefficient $\alpha$ was set to $1$ in all our simulations. A test has been performed to verify that assuming a value of 0.9 does not change the results significantly (see Section~\ref{ss:rubustness}).

\subsubsection{Modeling 2016~GE1-based parameters}
\label{sss:GE1_param}
These parameters are modelled according to their values determined for the asteroid 2016~GE1.
All these are assumed to be Gaussian distributed, with a mean value equal to the nominal estimated value and standard deviation equal to the $1\sigma$ uncertainty of the measurement. 
The value of the measured semi-major axis drift $(\text{d}a/\text{d}t)_{\text{m}}$ has been estimated through orbit determination, and it is reported in Table~\ref{tab:orb_param}.
The absolute magnitude $H$, although not explicitly present in Eq.~\eqref{eq:dadt_i}, is needed for the model by \citet{2021A&A...647A..61F} in order to construct a population-based distribution of the diameter $D$ and of the density $\rho$ (see Sec.~\ref{ss:popm}). The measured absolute magnitude of 2016~GE1 is $H = 26.7$, and we used the uncertainty of 0.5 given by \texttt{OrbFit} after the convergence of orbital fit. 
%

The rotation period reported in the Asteroid Light Curve Database \citep[LCDB,][]{warner-etal_2009} is $P = 0.009438$ h, which corresponds to about 34 s. The lightcurve was obtained with an exposure time of 10 s \citep{2016MPBu...43..240W}, and the quality code $U$ reported in LCDB corresponds to 2, implying an uncertainty of about 30\%. Therefore, we used a $\sigma = 0.3 \times P$ for this parameter. In fact, the $U=2$ code flag might also imply that the period could be wrong by an integer multiple. However, the short period mentioned above was recently confirmed also by \citet{2022DPS....5452306G}, which gives us some confidence that it is accurate. Therefore, despite this limitation, 2016~GE1 is very likely the NEA with the shortest rotation period for which the Yarkovsky effect has been determined so far. Nevertheless, in Section~\ref{s:results} we analysed how the estimated thermal inertia changes with rotation period, and the limitations this might impose on the results.

\subsubsection{Modeling population-based parameters} 
\label{ss:popm}
The density $\rho$ and the diameter $D$ of 2016~GE1 are both unknown. Therefore, we use the population-based distribution model by \citet{2021A&A...647A..61F}.
The model combines the NEA orbital distribution by \citet{granvik-etal_2018} and the NEA albedo
distribution by \citet{morbidelli-etal_2020}, and provides a bi-variate 
distribution of the couple $(\rho, D)$. 
Given the orbital elements $a, e, i$ and the absolute magnitude $H$ of an object, the population model 
by \citet{granvik-etal_2018} provides the probability for an NEA to originate from each main-belt 
source region, that are: the $\nu_6$ secular resonance, the 3:1, 5:2, and 2:1 mean-motion
resonances with Jupiter, the Hungaria region, the Phocaea region, and the Jupiter-Family Comets (JFC). 
Note that the absolute magnitude value of 2016~GE1 is beyond the limit of $H=25$ of
validity of the model by \citet{granvik-etal_2018}, therefore the source-route
probabilities (reported in Table~\ref{table:routes}) are extracted by a linear
interpolation.
These probabilities are then combined with the NEA albedo distribution by
\citet{morbidelli-etal_2020}, and a PDF $p_{p_V}$ for the albedo $p_V$ is determined first
\citep[see][for details]{2021A&A...647A..61F}. 

To obtain a distribution of $(\rho, D)$, we sample the albedo according to its PDF
$p_{p_V}$. For each point of the
sample, we produce a value of the diameter $D$ by using the conversion formula
\citep[see e.g.][]{bowell-etal_1989, 2007Icar..190..250P}
\begin{equation}
   D = \frac{1329 \text{ km}}{\sqrt{p_V}}10^{-H/5}.
   \label{eq:pv2dia}
\end{equation}
The same albedo value is used to generate a value of the density $\rho$. To this end, we 
divide the albedo into three categories, and associate an asteroid complex to each of them:
$p_V \leq 0.1$ is associated the C-complex, $0.1 < p_V \leq
0.3$ is associated with the S-complex, and $p_V > 0.3$ to the
X-complex.\footnote{We recall that the X-complex is degenerate in terms of albedo, containing both low- and high-albedo objects. It seems, however, to be also degenerate in terms of density, including not only asteroids of high density, but also of low density \citep[see Figure 5 in][]{2023A&A...671A.151B}. Moreover, the low-density X-complex objects are typically those of the P-type \citep{2013ApJ...762...56U}, which are less likely to be present among the NEAs. Therefore, X-complex asteroids in the near-Earth region are typically those of high albedo \citep[][Table 4]{2011AJ....142...85T}. For these reasons, we found it reasonable to assume here that high-density X-type asteroids are also of high albedo.} A value of the density $\rho$ is then generated according to the class in which the selected albedo value falls in. The density of each group is assumed to be log-normal distributed, with the average and the standard deviation listed in Table~\ref{tab:astDensities}.
\begin{table}[!bt]
    \caption{Average density and the standard deviation of the three asteroid
    complexes, as used in \citet{2021A&A...647A..61F}.}
    \centering
    \begin{tabular}{cc}
    \hline
    \hline
         Complex & Density (kg m$^{-3}$) \\
         \hline
            C    & 1200 $\pm$ 300 \\
            S    & 2720 $\pm$ 540 \\
            X    & 2350 $\pm$ 520 \\
         \hline
    \end{tabular}
    \label{tab:astDensities}
\end{table}
\begin{table}[!ht]
   \caption{Source-region probabilities of 2016~GE1, taken from \citet{granvik-etal_2018}.}
   \centering
   \begin{tabular}{cc} 
      \hline
      \hline
      Source region & Probability \\ [0.5ex] 
      \hline
      $\nu_6$  & 0.7958 $\pm$ 0.0475 \\
      3:1      & 0.1123 $\pm$ 0.0262 \\
      5:2      & 0.0002 $\pm$ 0.0039 \\
      Hungaria & 0.0911 $\pm$ 0.0325 \\ 
      Phocaea  & 0.0000 $\pm$ 0.0017 \\
      2:1      & 0.0002 $\pm$ 0.0007 \\
      JFC      & 0.0003 $\pm$ 0.0001 \\
      \hline
 \end{tabular}
 \label{table:routes}
\end{table}

\begin{figure}[!ht]
    \centering
    \includegraphics[width=0.51\textwidth]{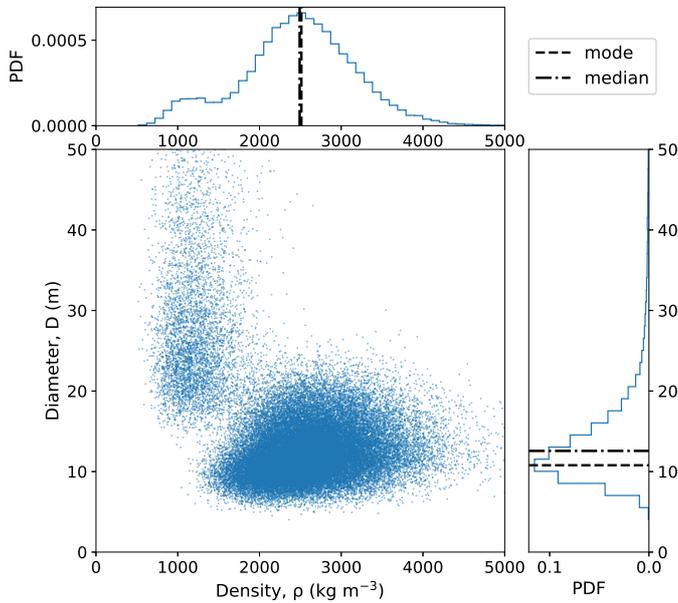}
   \caption{The input density $\rho$ versus diameter $D$ distribution for 2016~GE1. The blue histograms at the top and right show the marginal distributions of $\rho$ and $D$, respectively.}
   \label{fig:2016GE1_rho_D}
\end{figure}

Figure~\ref{fig:2016GE1_rho_D} shows the joined distribution of $(\rho, D)$ obtained for 2016~GE1, where the 
correlation given by the albedo can be seen by the fact that smaller size is associated with larger density (moderate and large albedo cases), while the larger size is associated to smaller density (low albedo case). The marginal PDFs of $\rho$ and $D$, i.e. the distribution of the set of all the possible $\rho$ (resp. $D$) values alone, are also shown in Fig.~\ref{fig:2016GE1_rho_D}. 
The most likely value and the median of the density are almost the same, and
they are about 2490 kg m$^{-3}$. The most likely value of the diameter is 12 m, while
the median value is 14 m.

The obliquity $\gamma$ is also unknown. Therefore we assume it to be distributed according to
the NEA obliquity distribution determined by \citet{2017A&A...608A..61T}. This distribution 
has a 2:1 ratio between retrograde and prograde rotators. However, the model
always rejects values of $\gamma$ that are not compatible with the sign of the measured Yarkovsky drift 
because, in this case, solutions to Eq.~\eqref{eq:yarkoInvertFormula} can not be found.

\section{Basic model testing and verification}
\label{sec:testing}

An essential step in the application of any new method is its testing and verification. To this purpose, we used asteroid (101955) Bennu, the target of NASA's OSIRIS-REx sample return mission, for which all relevant input and output parameters are well constrained, allowing us to test how different assumptions or unknown parameters affect the results.

\begin{table*}[]
    \caption{Orbital and physical parameters of asteroid (101955) Bennu.}
    \centering
    \begin{tabular}{lcc}
         \hline
    Parameter & Value & Reference \\
         \hline
    Semi-major axis, $a$   & 1.1259963567892803  $\pm$ 1.718E-10 au   &  NEOCC - Epoch 60000.0000 MJD \\
    Eccentricity, $e$  &  0.203719194929926  $\pm$ 2.045E-8   & NEOCC - Epoch 60000.0000 MJD \\
    Semi-major axis drift, $da/dt$  & $-$284$\pm0.2$ m yr$^{-1}$ & \citet{2021Icar..36914594F} \\
    \hline
    Absolute magnitude, $H$  & 20.21 $\pm$ 0.03 & \citet{2019NatCo..10.1291H} \\
    Radius, $r$  & 242.22 $\pm$ 0.15 m & \citet{2020SciA....6.3649D} \\
    Bulk density, $\rho$  & 1194$\pm$3 kg m$^{-3}$  & \citet{2020SciA....6.3649D} \\
    Obliquity, $\gamma$  & 177.6 $\pm$ 0.11 degrees & \citet{lauretta-etal_2019} \\
    Rotation period, $P$  & 4.2960015 $\pm$ 0.0000018 hours & \citet{2020SciA....6.3649D} \\
    Albedo, $p_V$   & 0.044 $\pm$ 0.002 & \citet{lauretta-etal_2019} \\
    Emissivity, $\epsilon$   & 0.984 &  \\
    \hline
    Thermal inertia,  $\Gamma$   & 310$\pm$70 J m$^{-2}$ K$^{-1}$ s$^{-1/2}$ & \citet{2014Icar..234...17E} \\
    Thermal inertia,  $\Gamma$   & 350$\pm$20 J m$^{-2}$ K$^{-1}$ s$^{-1/2}$ & \citet{2019NatAs...3..341D} \\
    Thermal inertia,  $\Gamma$   & 300$\pm$30 J m$^{-2}$ K$^{-1}$ s$^{-1/2}$ & \citet[][OTES]{2020SciA....6.3699R} \\
    Thermal inertia,  $\Gamma$   & 320$\pm$30 J m$^{-2}$ K$^{-1}$ s$^{-1/2}$ & \citet[][OVIRS]{2020SciA....6.3699R} \\
         \hline
    \end{tabular}
    \label{tab:bennu}
\end{table*}

The values of orbital and physical parameters of Bennu used in our model are given in Table~\ref{tab:bennu}.
We note that there are different estimations of Bennu's thermal inertia. Based on the Spitzer Space Telescope measurements of the Bennu's thermal emission, \citet{2014Icar..234...17E} derived thermal inertia of $310 \pm 70$ J m$^{-2}$ K$^{-1}$ s$^{-1/2}$. A global-average thermal inertia of $350\pm20$ J m$^{-2}$ K$^{-1}$ s$^{-1/2}$ was estimated by \citet[][]{2019NatAs...3..341D}, using the OSIRIS-REx approach-phase thermal emission light curves and the encounter-based shape model. More recently, \citet{2020SciA....6.3699R} analysed the data from the OSIRIS-REx Thermal Emission Spectrometer (OTES) and the OSIRIS-REx Visible and InfraRed Spectrometer (OVIRS), and derived mean thermal inertia values of $300 \pm 30$ and $320 \pm 30$ J m$^{-2}$ K$^{-1}$ s$^{-1/2}$ from OTES and OVIRS, respectively. The authors also found spatial variations in thermal inertia, with larger values closer to the equator.    

In principle, Bennu's thermal inertia values from the literature are very similar and generally consistent. We will primarily compare our test results with values from \citet{2020SciA....6.3699R}, but conclusions remain broadly the same if the other values are used for comparison.   

\begin{table}[]
    \caption{Summary of the model tests on asteroid Bennu. See also Fig.~\ref{fig:Bennu_TI}.}
    \centering
    \begin{tabular}{llcc}
         \hline
    Test & Parameters used & ~~~~~~Thermal & Inertia ~~~~~~~~ \\
     &   & left peak & right peak \\
         \hline
     01 & $a,e,H,P$ & $\Gamma = 97 \pm 39$  & $\Gamma = 592 \pm 257$  \\
     02 & $a,e,H,P, p_V,D$ & $\Gamma = 108 \pm 37$ & $\Gamma = 505 \pm 189$ \\
     03 & $a,e,H,P, p_V,D, \rho$ & $\Gamma = 139 \pm 26$ &  $\Gamma = 346 \pm 64$ \\
     04 &  $a,e,H,P, p_V,D, \rho, \gamma $ &  $\Gamma = 122 \pm 22$  & $\Gamma = 387 \pm 65$ \\
         \hline
    \end{tabular}
    \label{tab:bennu_tests}
\end{table}

\begin{figure*}
   \centering
   \includegraphics[width=17cm]{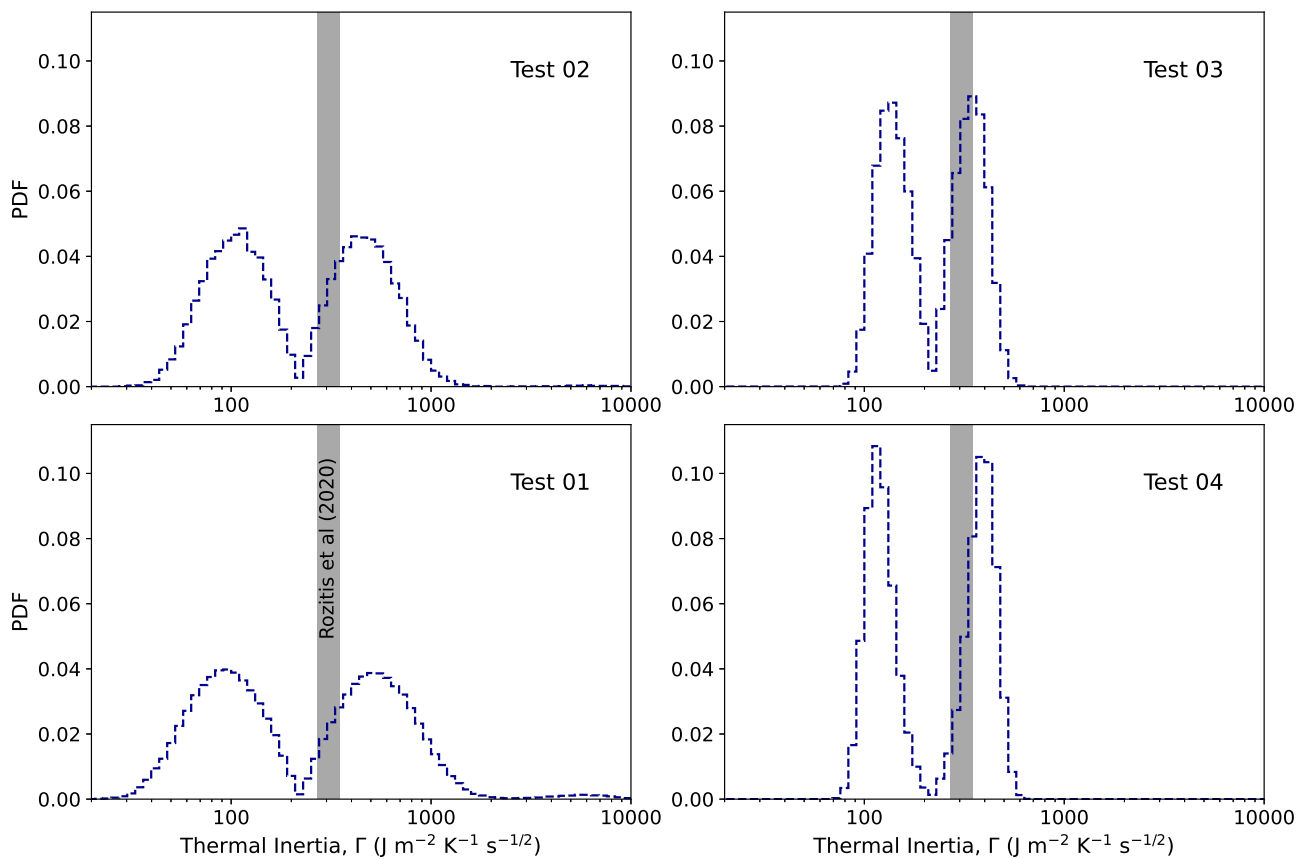}
 \caption{Monte Carlo model tests on asteroid Bennu. The panels show the distributions of thermal inertia solutions for different input parameters assumed to be known. The number of known parameters increases clockwise, starting from the bottom-left panel. For additional details on the test, see Table~\ref{tab:bennu_tests}. The grey area in each panel marks the interval of the thermal inertia of Bennu as estimated by \citet{2020SciA....6.3699R}.  }
   \label{fig:Bennu_TI}
\end{figure*}

To test our model, we adopted the following strategy. Since for a typical asteroid we know much less data than for Bennu, we started with just basic properties. Then, in each subsequent step, we added some new parameters with better constraints. In this way, we expect to show that the model provides meaningful results even with only basic knowledge about the object, but that the results become more accurate when additional knowledge is available. Therefore, we performed four tests. In the first test, we assumed that only basic information about the object are available: the orbit (semi-major axis and eccentricity), absolute magnitude and rotation period. In the second test, albedo and diameter are included as well. The third test also includes Bennu's density, while in the final fourth test, we added the obliquity. The information about the test are summarized in Table~\ref{tab:bennu_tests}, while the results are also shown in Fig.~\ref{fig:Bennu_TI}.

Before we discuss the results obtained for Bennu, let us recall that the thermal inertia solution is typically bi-modal, resulting in two prominent peaks in the obtained distribution. While both solutions are possible in principle, based on the empirical understanding, we favour the right peaks. Therefore, in what follows, we will refer to the right peaks in the distribution of thermal inertia as our nominal results. In some cases, depending on the exact goal, we may speak about maximum (or minimum) values of the thermal inertia, limiting the results from one side rather than providing the most likely value. 

The result of the first test ($\Gamma = 592 \pm 257$ J m$^{-2}$ K$^{-1}$ s$^{-1/2}$), which is significantly based on the NEOs' population models, is skewed towards higher values and the corresponding uncertainty is large. This shows that without any prior knowledge of the physical parameters, the results obtained for a single asteroid are unreliable in a general case. However, despite the limitations of the large uncertainty, the result is still statistically compatible with the values found in the literature, except the one derived by \citet{2020SciA....6.3699R} from the OTES instrument (see Tables~\ref{tab:bennu}. and \ref{tab:bennu_tests}). Adding knowledge about the albedo and diameter in the second test resulted in a somewhat improved result and reduced uncertainty ($\Gamma = 505 \pm 189$ J m$^{-2}$ K$^{-1}$ s$^{-1/2}$). Including knowledge of density in the third test changed the situation significantly. The obtained thermal inertia of $\Gamma = 346 \pm 64$ J m$^{-2}$ K$^{-1}$ s$^{-1/2}$ is now fully in line with the high-accuracy measurments. Interestingly, when the information about the obliquity is added in the fourth test, the result is moved further from the referent value, though still plausible.

Based on the tests presented above, we conclude that our Monte Carlo-based model for thermal inertia determination could be useful even when only basic information about an object are known. In this case, however, the uncertainty of the result could be large, and the model may not be fully appropriate for individual objects. Noticeable exceptions from this are rapidly rotating objects (see Section~\ref{ss:rubustness}). 

On the other hand, if most of the input parameters are sufficiently well-known, the model provides accurate and reliable results even in the case of an individual object. 

\section{Estimated thermal characteristics of 2016~GE1}
\label{s:results}
We performed the MC estimation of the thermal conductivity $K$ for four fixed heat capacity values, namely $C=600,800,1000$, and $1200$ J kg$^{-1}$ K$^{-1}$. A random sample of one million combinations of the input parameters was used.
Thermal conductivity solutions were searched in the range between $10^{-8}$ W m$^{-1}$ K$^{-1}$ and $500$ W m$^{-1}$ K$^{-1}$, that we believe to be more than appropriate considering the known variety of materials composing asteroids.  
For each $K$ solution, we also computed the corresponding thermal inertia as
\begin{equation}
   \Gamma = \sqrt{\rho K C}.
   \label{eq:K2G}
\end{equation}
Moreover, we run the estimation of $K$ with the semi-major axis drift solution given by the JPL SBDB first, and then run once again with the solution we obtained with \texttt{OrbFit}. 

\begin{figure*}
   \centering
   \includegraphics[width=17cm]{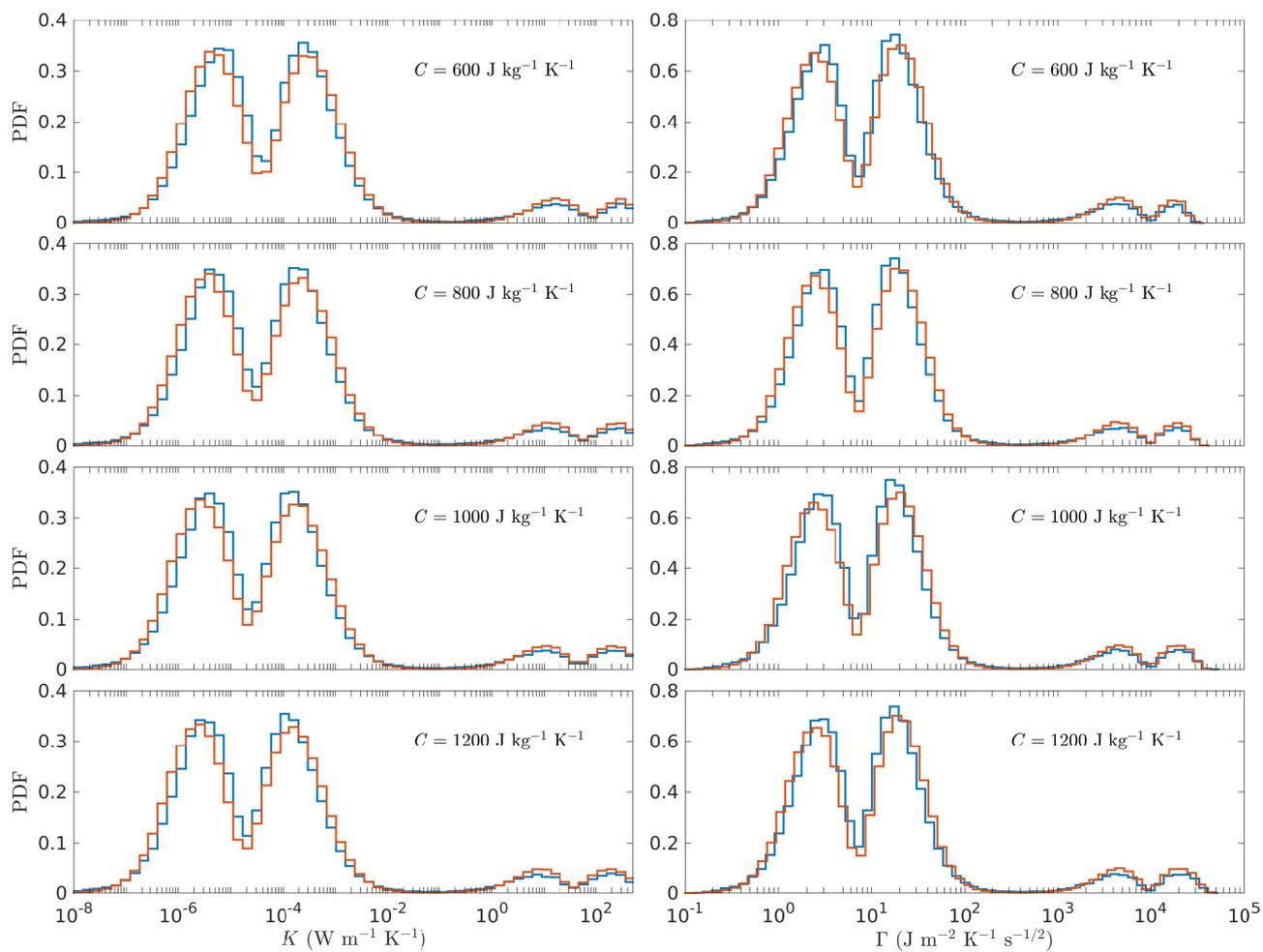}
   \caption{Distributions of the thermal parameters of 2016~GE1, for different values of
   heat capacity $C$. The first column shows the distributions of the thermal conductivity
$K$, while the second column the distributions of the thermal inertia $\Gamma$. Blue histograms are the results obtained by using the 
orbital solution provided by the JPL SBDB, while red histograms are the results obtained by using our solution obtained with \texttt{OrbFit}.}
   \label{fig:2016GE1_all}
\end{figure*}

Figure~\ref{fig:2016GE1_all} shows the distributions of $K$ and $\Gamma$
obtained for the different values of heat capacity $C$. Blue histograms refer to the results obtained with the JPL SBDB orbital solution, while red histograms to those obtained with our orbital fit.

All distributions have the same properties. For a thermal conductivity smaller than $\sim$10$^{-3}$ W m$^{-1}$ K$^{-1}$ (thermal inertia smaller than $\sim$100 J m$^{-2}$ K$^{-1}$ s$^{-1/2}$), two peaks always occur with high probability. Values for $K$ between $\sim$0.01 and $\sim$1 W m$^{-1}$ K$^{-1}$ and for $\Gamma$ between $\sim$100 and $\sim$1000 J m$^{-2}$ K$^{-1}$ s$^{-1/2}$ are extremely unlikely. Finally, for values $K \gtrsim 1$ W m$^{-1}$ K$^{-1}$ ($\Gamma \gtrsim 1000$ J m$^{-2}$ K$^{-1}$ s$^{-1/2}$), there is a low probability tail in which two more peaks occur.

The distributions are almost independent of the heat capacity $C$. Therefore, we always refer to the results for $C=600$ J kg$^{-1}$ K$^{-1}$ in the following, unless explicitly stated otherwise.
We performed the Kolmogorov-Smirnov test to check whether the distributions obtained for the two solutions for the semi-major axis drift are identical. The null hypothesis was rejected with a significance level of 5\%. Nevertheless, the two distributions are very similar in each of the cases shown in Fig.~\ref{fig:2016GE1_all}, and the results show only minor differences.

The two high probability peaks at low thermal conductivity appear at
\[
\sim 3-6 \times 10^{-6} \text{ W m$^{-1}$ K$^{-1}$} 
\]
and at 
\[ 
\sim 1-2\times 10^{-4} \text{ W m$^{-1}$ K$^{-1}$},
\] 
while the corresponding peaks in thermal inertia are at 
\[
3 \text{ J m$^{-2}$ K$^{-1}$ s$^{-1/2}$}
\]
and at
\[ 
18 \text{ J m$^{-2}$ K$^{-1}$ s$^{-1/2}$}.
\]

\begin{figure}
    \centering
    \includegraphics[width=0.48\textwidth]{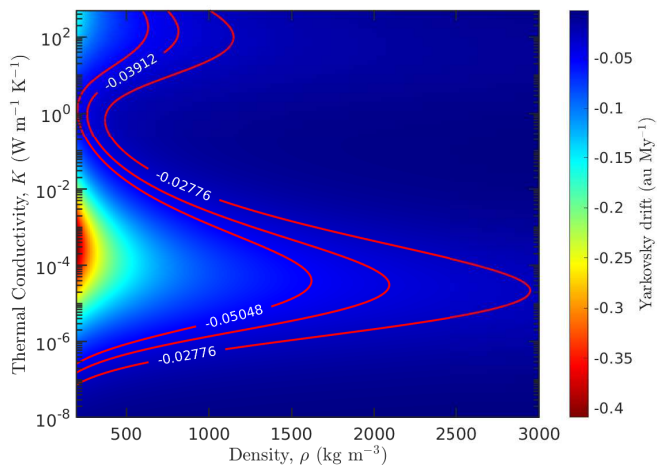}
    \caption{Estimated semi-major axis drift for 2016~GE1, obtained for $D =9.8$ m, $C = 800$ J kg$^{-1}$ K$^{-1}$, $\gamma = 133.77^\circ$, and $P = 0.009438$ h. The red level curves represent the Yarkovsky drift measured from astrometry with \texttt{OrbFit}, and the corresponding $1\sigma$-uncertainty.}
    \label{fig:yarko_grid}
\end{figure}

Note that these peaks are located at extremely low $K$, and they are almost an order of magnitude smaller than those obtained for 2011~PT \citep[][see also Fig.~\ref{fig:2011~PT}]{2021A&A...647A..61F}.
On the other hand, the two low probability peaks at high thermal conductivity are at $\sim$10 and $\sim$250 W m$^{-1}$ K$^{-1}$, corresponding to the thermal inertia of $\sim$4~450 and $\sim$18~000 J m$^{-2}$ K$^{-1}$ s$^{-1/2}$, respectively. The presence of these two low probability peaks is due to the fact that the measured vs. predicted Yarkovsky drift equation in Eq.~\eqref{eq:yarkoInvertFormula} has either 3 or 4 different thermal conductivity solutions for certain combinations of input parameters. To show an example of this behaviour, we computed the Yarkovsky drift on a grid in density $\rho$ and thermal conductivity $K$, by fixing the other parameters to $D = 9.8$ m, $C = 800$ J kg$^{-1}$ K$^{-1}$, $\gamma = 133.77^\circ$, and $P = 0.009438$ h. Figure~\ref{fig:yarko_grid} shows the contour plot of the computed semi-major axis drift. The levels corresponding to the $\text{d}a/\text{d}t$ solution computed with \texttt{OrbFit}, together with those at 1$\sigma$-uncertainty, are highlighted in red. From this figure, it can be appreciated that the measured vs. predicted Yarkovsky drift equation has either 3 or 4 solutions in the interval $K \in [10^{-8}, \ 500]$ W m$^{-1}$ K$^{-1}$, for certain values of the density $\rho$.

To give quantitative constraints of the thermal parameters, we fit the thermal inertia distributions by using the kernel density estimation, 
and then we computed the probability
\begin{equation}
   \begin{split}
   P_1 & = P(\Gamma < 100  \text{ J m$^{-2}$ K$^{-1}$ s$^{-1/2}$}), \\
   P_2 & = P(\Gamma > 1000 \text{ J m$^{-2}$ K$^{-1}$ s$^{-1/2}$}).
   \end{split}
   \label{eq:P1P2}
\end{equation}
The probability $P_1$ was always $\sim$0.92, with negligible differences between the two solutions of the semi-major axis drift. On the other hand, the probability $P_2$ was always $\sim$0.08, implying that solutions with high thermal inertia are unlikely. Therefore, the results show that the thermal inertia of 2016~GE1 is probably very low, which is unexpected for the extremely fast-rotating asteroid. 

\subsection{Robustness of the results}
\label{ss:rubustness}

The results presented suggest that 2016~GE1 has very low thermal inertia. In this subsection, we discuss how reliable such a conclusion is. The presented Monte Carlo-based model for asteroid thermal inertia estimations generally depends on a set of input parameters. As demonstrated in Section~\ref{sec:testing}, when these parameters are well known (or at least most of them), the model could provide good results for individual objects.  

On the other hand, when only basic information such as orbit, absolute magnitude, and Yarkovsky drift are known, the model relies on population-based models of input parameters. In this case, the model could still provide useful results to model the thermal inertia of a population of asteroids. However, the results for an individual object are uncertain and generally unreliable. The situation with the 2016 GE1 is very similar to that case, except that the rotation period is known, in addition to the orbit and absolute magnitude. \textit{So, why then the obtained results should be considered reliable? }  

The 2016 GE1 is a rapid rotator with a rotation period of only about 34 s. At the same time, a significant $A_2$ acceleration associated with the Yarkovsky effect has been measured. A temperature gradient across the surface must be present for this effect to work. However, in the case of such a rapid rotation, the temperature gradient could exist only in the case of low thermal inertia. As a result, the rotation period strongly constrains the range of acceptable thermal inertia. In the case of 2016 GE1, only about 16\% of input parameter combinations are accepted in our simulations as possible. This is largely due to solid constraints from the rotation period.    

\begin{figure*}
    \centering
\includegraphics[width=17cm]{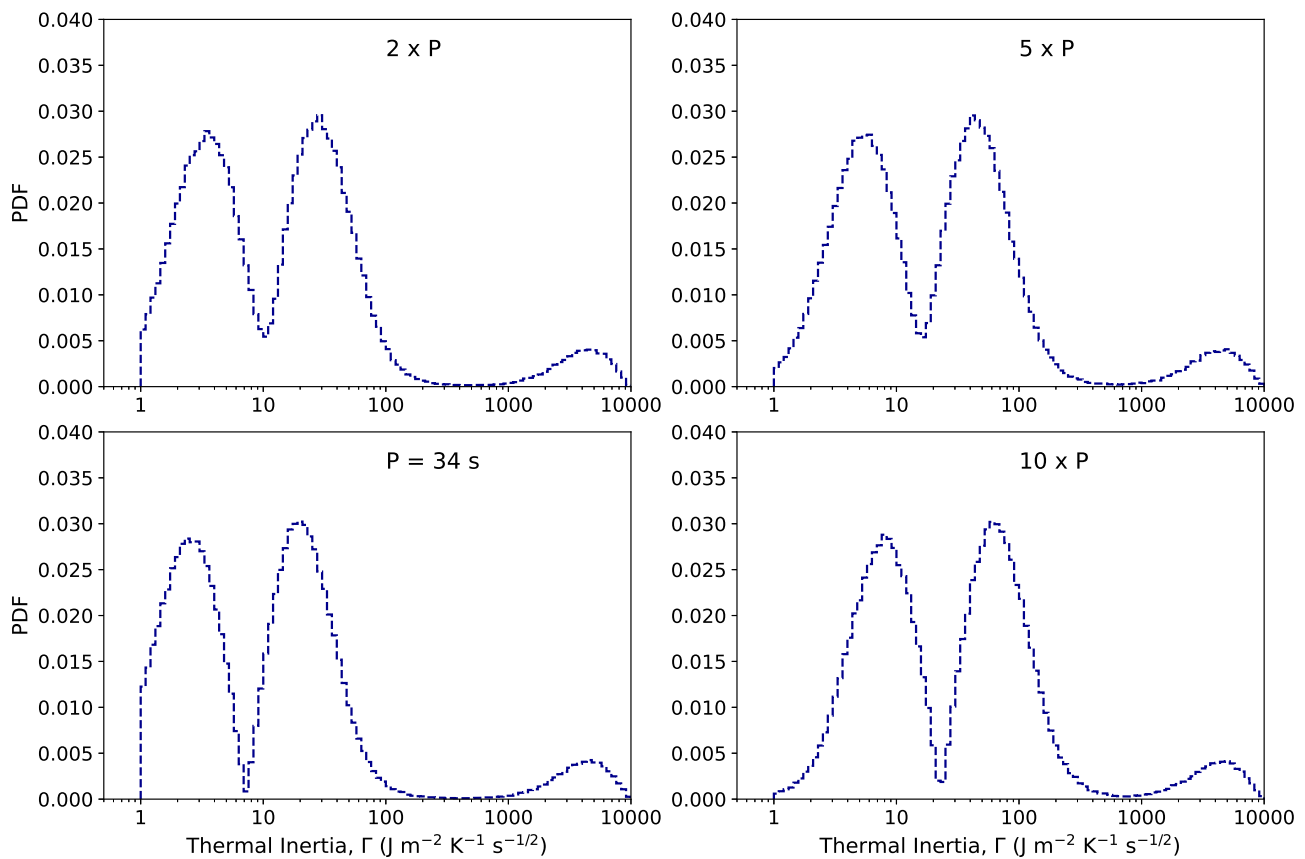}
\caption{The dependence of the 2016 GE1's thermal inertia estimation on the rotation period. The assumed period increases clockwise from the bottom-left panel.The results are shown for the nominal period solution of 34 seconds, as well as for 2, 5 and 10$\times$ longer periods, as indicated in the plots.}
    \label{fig:GE1_TI_rot}
\end{figure*}

As explained in Section~\ref{sss:GE1_param}, despite some uncertainties, we believe the short rotation period of 2016 GE1 is determined reliably enough. Nevertheless, we tested what would happen if the rotation period is 2, 5 and 10 times longer than the measured one. In Figure~\ref{fig:GE1_TI_rot}, we show how the resulting thermal inertia of 2016 GE1 depends on the rotation period. Increasing the rotation period shifts the distribution of thermal inertial to the right, i.e. towards larger values. Still, even for a $10 \times$ longer rotation period, more than 90\% of the solution suggests thermal inertia below 300 J m$^{-2}$ K$^{-1}$ s$^{-1/2}$, which can be considered low.   

There are indeed small peaks in the TI distribution associated with values of $\Gamma > 1000$ J m$^{-2}$ K$^{-1}$ s$^{-1/2}$. These values are incredibly high to the point of physical implausibility, and other studies have found only very few objects with such high thermal inertia estimates \citep{2022PSJ.....3...56H}. We, therefore, discarded them as highly improbable.  

\begin{figure}
    \centering
\includegraphics[width=0.45\textwidth]{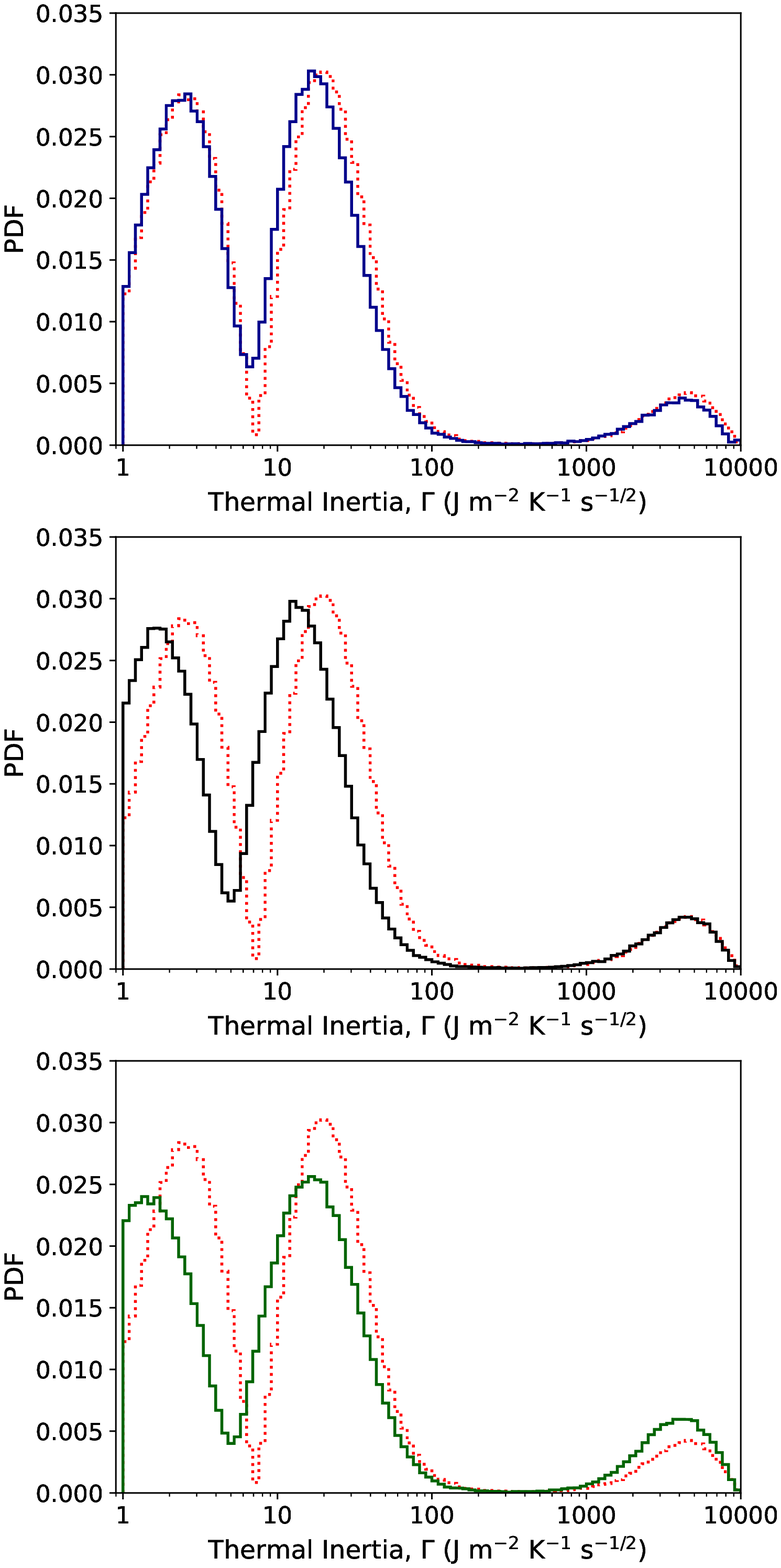}
\caption{The dependence of the resulting thermal inertia estimation on the input parameters - comparison with the nominal results. As a guide, in each plot, the results of our nominal estimation are shown as a red-dotted histogram. The top panel shows the obtained thermal inertia distribution for absorption coefficient $\alpha$ set to 0.9 (blue histogram). The middle panel shows the results obtained assuming an emissivity $\varepsilon$ of 0.9 (black histogram). The bottom panel shows how the thermal inertia distribution changes when a uniform input distribution of density $\rho$ is assumed (green histogram). }
    \label{fig:GE1_TI_par}
\end{figure}

We also tested how the assumed values of the emissivity $\varepsilon$ and the absorption coefficient $\alpha$ may affect the results. We found that for a reasonable range of these parameters, the changes in results are small and cannot affect general conclusions about the low thermal inertia of 2016~GE1 (see Figure~\ref{fig:GE1_TI_par}). Additionally, we investigated how the assumption of entirely random input distribution of the density $\rho$ in the range 1000-3500 kg m$^{-3}$ would change the result. Again, we found that the resulting thermal inertia values are not much different. Moreover, they are even shifted towards lower values (bottom panel of Figure~\ref{fig:GE1_TI_par}). 

\begin{figure}[!ht]
    \centering
    \includegraphics[width=0.47\textwidth]{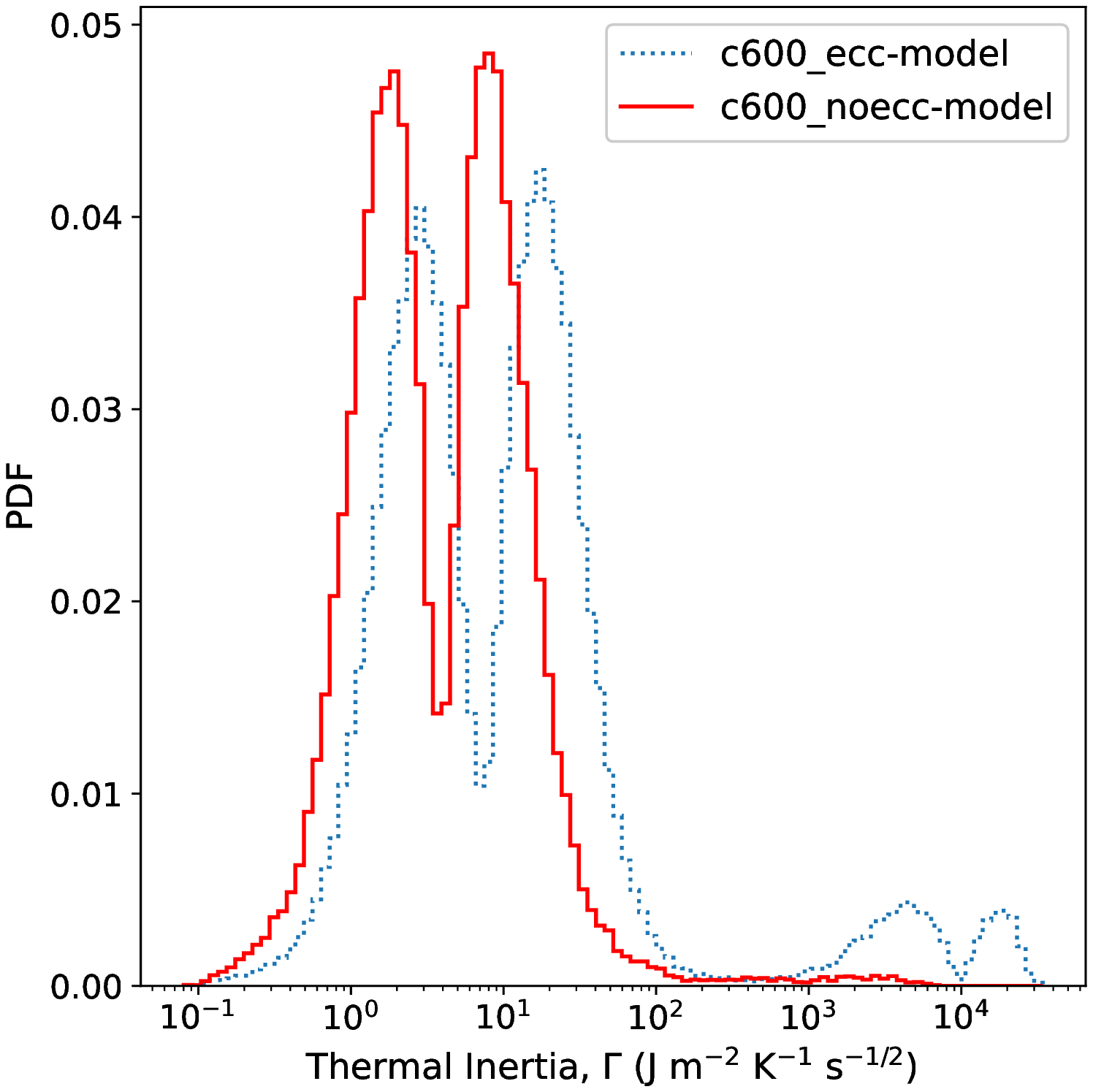}
   \caption{The resulting thermal inertial distribution $\Gamma$ with eccentric (blue histogram) and circular (red histogram) Yarkovsky model. In both cases, the heat capacity was set to $C = 600$ J kg$^{-1}$ K$^{-1}$. See text for more details.}
   \label{fig:2016GE1_ecc_noecc}
\end{figure}

To conclude this part, let us show how the results change when the eccentric and circular Yarkovsky models are used. As already mentioned in Section~\ref{s:yarko_model}, the Yarkovsky model based on the assumption of a circular orbit may not be suitable due to the large eccentricity of the orbit. In Figure~\ref{fig:2016GE1_ecc_noecc}, we show how the results differ for two Yarkovsky models when all other parameters are equal. The thermal inertia solutions in the case of the circular model favour even smaller values. Two peaks at low $\Gamma$ are slightly shifted to lower values, while two smaller peaks found in the eccentric model at high values of thermal inertia disappear in the circular model. This demonstrates that the circular model is not fully suitable for NEAs that are in moderately to highly eccentric orbits. However, in the case of 2016~GE1, the results seem to be mainly driven by its rapid rotation. They, therefore, do not vary much for different input parameters or the Yarkovsky model.

With this in mind, we concluded that our estimate of the 2016 GE1 low thermal inertia is robust, even though we cannot very accurately estimate its nominal value. Nevertheless, we showed that thermal inertia of GE1 cannot exceed $\Gamma = 300$ J m$^{-2}$ K$^{-1}$ s$^{-1/2}$. The real thermal inertia is likely even smaller, with a probability of $>90$\% to be below 100 J m$^{-2}$ K$^{-1}$ s$^{-1/2}$.

\subsection{Posterior distribution of the input parameters}
In addition to calculating the thermal inertia, we kept track of all combinations of input parameters for which at least a single solution of Eq.~\eqref{eq:yarkoInvertFormula} was found. If no solutions are found, the measured semi-major axis drift cannot be achieved for the chosen combination, and therefore these determined values of the physical parameters are not representative for 2016~GE1.

\begin{figure}
    \centering
\includegraphics[width=0.48\textwidth]{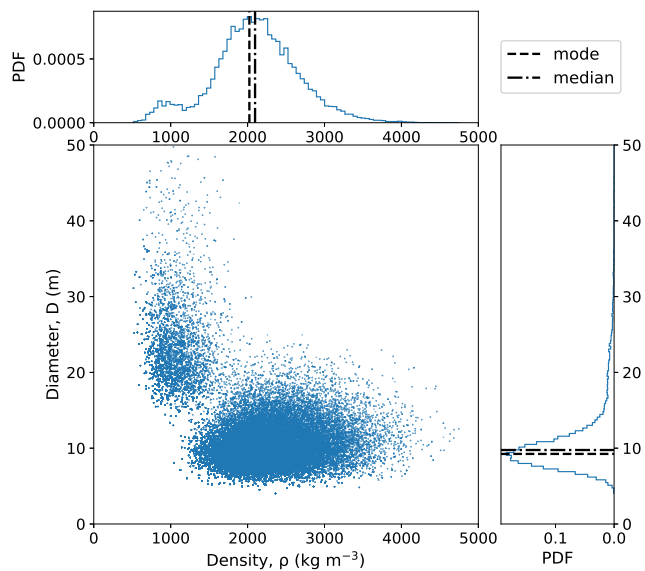}
\caption{The posterior (resulting) distribution of the parameters of GE1. The main plot shows the density $\rho$ against the diameter distribution $D$. The upper and right histograms show the marginal distributions of $\rho$ and $D$, respectively. The results were obtained using the eccentric Yarkovsky model and the heat capacity of $C = 600$ J kg$^{-1}$ K$^{-1}$. The dashed and dash-dotted lines mark the modes and medians of the distributions, respectively.}
    \label{fig:rho_D_out}
\end{figure}

Figure~\ref{fig:rho_D_out} shows the $2$-D distribution of $(\rho, D)$ and their marginal distributions obtained for the simulations using the semi-major axis solution provided by the JPL SBDB and $C = 600$ J kg$^{-1}$ K$^{-1}$. The results in all other cases are similar.
The correlations low density$-$large diameter and high density$-$small diameter are still present in the output distributions, as shown in the top panel of Fig.~\ref{fig:rho_D_out}.
The median value and the most likely diameter are 9.8 and 9.2 meters, respectively, which are both smaller than the values obtained for the input distribution in Fig.~\ref{fig:2016GE1_rho_D}. This means that the solutions of Eq.~\eqref{eq:yarkoInvertFormula} are more likely to be found for small diameters. From this distribution, we determined that 2016~GE1 has a diameter ranging from 5 m to 15 m with probability $\sim$0.87 and is smaller than 20 m with probability $\sim$0.94.

The density distribution peaks at about 2020 kg m$^{-3}$, with a median value of $\sim$2100 kg m$^{-3}$, both lower than the corresponding values of the input distribution. The probability of 2016~GE1 having a density of less than 1200 kg m$^{-3}$ or more than 3000 kg m$^{-3}$ is small, $\sim$0.06 in each case. We note, however, that the density distribution is two-peaked, as expected, with each peak generally corresponding to one of two main taxonomic complexes of C- and S-type asteroids. Therefore, if asteroid 2016~GE1 is found to belong to the C-type, the larger values of the density should be rejected. Additionally, in case 2016~GE1 is an X-complex low-albedo but a high-density object, the assumed correlation between diameter and density would be broken, and the analysis of the posterior parameter distribution would be meaningless. We consider, however, this scenario unlikely. 

\section{Discussion}
\label{s:dis}
The values of $\Gamma < 100$ J m$^{-2}$ K$^{-1}$ s$^{-1/2}$ are obtained with a probability of $\sim$92 percent. Moreover, we have tested the stability of the result
with respect to various input parameters and verified their reliability. Therefore, the results presented above strongly suggest a very low thermal inertia of the rapidly rotating asteroid 2016~GE1. 

These thermal inertia values are not new for asteroids \citep[see e.g.][]{delbo-etal_2015}.
For example, the largest asteroids, such as (1)~Ceres or (4)~Vesta, have very low thermal inertia 
values of $\Gamma < 50$ J m$^{-2}$ K$^{-1}$ s$^{-1/2}$ as measured \citep{2012A&A...539A.154L,2020JGRE..12505733R}, and it is due to the fine regolith present at their surfaces, that is made possible by the relatively large gravity of these objects.

Another cause for low thermal inertia has been recently discovered during the OSIRIS-REx
and Hayabusa~2 missions, and it lies in the high micro-porosity of boulders.  The
ground-based estimated thermal inertia of $\Gamma = 310 \pm 70$ J m$^{-2}$ K$^{-1}$
s$^{-1/2}$ of Bennu \citep{2014Icar..234...17E} suggested a fine regolith-covered surface,
that was not found when OSIRIS-REx arrived at the asteroid \citep{lauretta-etal_2019}.
\citet{2020SciA....6.3699R} explained the measurements with the high micro-porosity of
Bennu's surface boulders, and suggested that this could be a characteristic property of
C-type asteroids. Furthermore, \citet{2021Natur.598...49C} found that the thermal inertia 
of Bennu’s rocks is positively correlated with the local surface abundance of fine regolith.
Similar results were obtained for Ryugu, where the global thermal inertia was estimated to
be $\Gamma = 225 \pm 45$ J m$^{-2}$ K$^{-1}$ s$^{-1/2}$ \citep{shimaki-etal_2020}, but no
fine regolith-covered surface was found \citep{watanabe-etal_2019}. In-situ images taken
by the Mobile Asteroid Surface Scout \citep[MASCOT;][]{2017SSRv..208..339H} lander did not
detect fine regolith on the boulders. Models of the temperature variations did not fit the
measurements when the regolith layer was taken into account, and the low thermal inertia was again
explained by the high porosity of the boulders \citep{grott-etal_2020}.

The spatially unresolved thermal inertia of comets is also very low, typically below $50$ J m$^{-2}$ K$^{-1}$ s$^{-1/2}$ \citep[e.g.][]{2018A&A...616A.122M,2019SSRv..215...29G}, although spatially resolved data indicate surface variations. In the case of comet 67P/Churyumov-Gerasimenko, for example, most values are in the range $10-170$ J m$^{-2}$ K$^{-1}$ s$^{-1/2}$. The thermal inertia of smooth terrains covered with deposits is lower (typically lower than $30$ J m$^{-2}$ K$^{-1}$ s$^{-1/2}$) than those of exposed consolidated terrains (typically larger than $110$ J m$^{-2}$ K$^{-1}$ s$^{-1/2}$) \citep{2019SSRv..215...29G}.

\textit{What are then the possible explanations for the 2016~GE1 and the most likely scenario?} Low thermal inertia is unexpected for super-fast rotating asteroids. On the one hand, this is because the fast rotation should clear out the surface of loose regolith. On the other hand, a large porosity often points towards a rubble-pile structure, which is also unexpected because the internal strength of such bodies might be too weak to maintain the integrity of the body under high rotational acceleration. Despite their improbability, however, none of the scenarios can be discarded, as recent models suggest that they are still possible \citep{sanchez-scheeres_2020,2021MNRAS.502.5277H}.
The integrity of the fast rotators could be maintained by a low yield stress of the order of 25 Pa. This level of yield stress can only be explained by assuming that the particles in asteroids are small enough (on the order of a few micrometres) to form a cement matrix (glue) between the larger particles (fragments) \citep{2021arXiv211015258P}.

Still, the most likely values for the thermal inertia found for 2016~GE1 by using the semi-analytical Yarkovsky model are significantly below $100$ J m$^{-2}$ K$^{-1}$ s$^{-1/2}$, smaller even than the thermal inertia of the boulders on Bennu and
Ryugu. 
For the single boulder on Ryugu investigated by the MASCOT lander, \citet{2022NatCo..13..364H} estimated the thermal inertia of 256$^{+4}_{-3}$ J m$^{-2}$ K$^{-1}$ s$^{-1/2}$, corresponding to an expected porosity of 46.7$^{+0.3}_{-0.4}$\%. 
This means that, if 2016~GE1 was a boulder, than it would be more porous than those
on Bennu or Ryugu.
To get a general idea of what porosity would explain the obtained values of thermal conductivity and inertia, we applied an empirical relationship between conductivity and porosity given by \citet{grott-etal_2020} \citep[see also][]{2016A&A...589A..41H} for meteorites. 
The employed empirical relation has its own limitations and, in particular, it is unreliable for large porosity.
Nevertheless, it can give a first indication of the degree of porosity in 2016 GE1 needed to explain its thermal properties.
We found that values of thermal inertia $\Gamma$ $\leq$ 20 J m$^{-2}$ K$^{-1}$ s$^{-1/2}$, compatible with two the most prominent peaks (see e.g. Fig.~\ref{fig:2016GE1_all}), are only feasible for porosity $\geq$~70\% and density $\rho \leq$ 1000 kg m$^{-3}$. Therefore a highly porous, Ryugu-boulder-like object is generally consistent with our result. If so, it also means that most of the density solutions should be discarded, and only those generally compatible with C-type asteroids can be considered.
This opens the possibility of small and super-fast rotators having a common origin as
anomalously low thermal inertia boulders similar to those found on Ryugu, which may be worth exploring in the future since they are the most similar to the primordial planetesimals \citep{2021NatAs...5..766S}.

On the other hand, \citet{sanchez-scheeres_2020} developed a model to study under what rotational conditions 
an asteroid can keep thin regolith on the surface, assuming that the asteroid has a
monolithic internal structure. The authors found that regolith can survive even at very
small rotation periods, especially in regions at high latitudes. Given the values of low thermal inertia, we found at 2016~GE1
which are the most consistent with the thermal inertia of dust-covered asteroids and the results of \citet{sanchez-scheeres_2020}
a plausible explanation for the low thermal inertia we found at 2016~GE1 is a dust layer.

In addition to the two explanations discussed above, another cause of low thermal inertia has emerged recently, namely the
\textit{cracked surface}. \citet{2023IJT....44...51I} analysed samples from asteroid Ryugu and found that the thermal inertia of the
samples is about 3.5 times larger than the observed thermal inertia of the asteroid Ryugu's surface.
The authors suggested that this difference in thermal inertia between mm- to cm-sized returned samples and 
boulders could be due to the presence of large-scale cracks caused by meteor impacts \citep[e.g.][]{2020Natur.587..205B} and thermal stresses \citep[][]{molaro-etal_2020,2022NatGe..15..453D} on a scale larger than several hundreds of micrometers in rocks and boulders on Ryugu. Therefore, the low thermal inertia of 2016 GE1 could also be due to the cracked surface.

\subsection{A population of a low thermal inertia super-fast rotating asteroids?}

\begin{figure}
    \centering
    \includegraphics[width=0.47\textwidth]{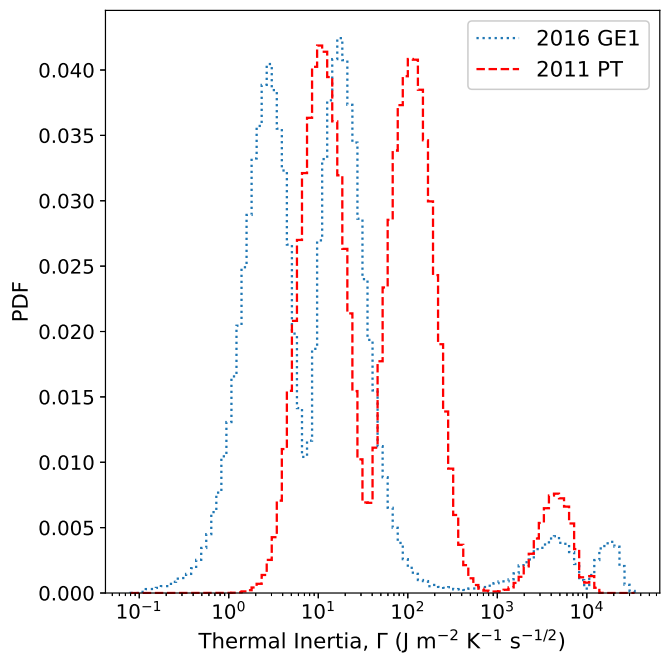}
    \caption{Comparative distributions of thermal inertia for the NEAs 2016~GE1 (blue dotted line) and 2011~PT (red dashed line). In both cases, the results were obtained with the eccentric Yarkovsky model and a heat capacity of $C=600$ J kg$^{-1}$ K$^{-1}$.}
    \label{fig:2011~PT}
\end{figure}

2016~GE1 is not the only super-fast rotating asteroid with low thermal inertia. As already mentioned,
the (499998) 2011~PT shares similar thermal inertia properties, and it has a diameter of about 35 m and rotates in only 11 minutes. The comparative results for the two objects are shown in Figure~\ref{fig:2011~PT}. 2016~GE1 is an extreme case of a small and super-fast rotator with extremely low thermal inertia, even lower than that found for 2011~PT. Nevertheless, the values of both objects are low and consistent with a dust-covered surface or a boulder with high micro-porosity. We note that the surface reflectivity properties of 2016~GE1 and (499998) 2011~PT are still unknown. Unfortunately, 2016~GE1 will not pass close to the Earth again in the next future, and therefore, it will not be observable due to its small size. On the contrary, 2011~PT will reach visual magnitudes smaller than 24 in June 2023, 2026, and 2029, and therefore, although still challenging, it may be observed with large-diameter telescopes. Any information in this regard would help us to better understand the nature of these unusual objects and unravel their mystery.

In addition, preliminary results obtained by \citet{2021EPSC...15..390P} on the NEA 1998~KY26, the target of the extended Hayabusa~2 mission, showed that it might also have a relatively low thermal inertia similar to that of 2011~PT. Further, by searching for objects with a determined $A_2$ in the JPL SBDB, many NEAs with $|d\text{a}/\text{d}t|>0.007$ au My$^{-1}$ (a value comparable with the semi-major axis drift of 2011~PT) can be found, and most of them have an absolute magnitude $H>24$. Even though information about their rotational state is not always known, the possibility that such small objects are fast rotators is still high, and therefore such fast semi-major axis drift would still be explained by low thermal inertia. It is also important to note that these $A_2$ measurements obtained by orbit determination are compatible with the Yarkovsky effect, except for a very small number of cases \citep{2023PSJ.....4...29F}. These facts open up the possibility for the existence of a new class of small super-fast rotating NEAs with low thermal inertia. 

\section{Summary and conclusions}
\label{s:conclusions}
In this work, we first performed a basic verification of our recently developed statistical MC method for determining asteroid thermal properties, showing that under the particular cases of super fast rotators with large measured $\text{d}a/\text{d}t$ due to the Yarkovsky effect we can constrain the surface thermal inertia. 

Then we used the model to constrain the thermal inertia of the small super-fast rotator 2016~GE1. This NEA has a diameter of less than 20 m, and its rotation period was estimated to be 34 seconds. We showed that thermal inertia of GE1 cannot exceed $\Gamma$ = 300 J m$^{-2}$ K$^{-1}$ s$^{-1/2}$. The real thermal inertia is likely even smaller, with a probability of >90\% to be below 100 J m$^{-2}$ K$^{-1}$ s$^{-1/2}$. The extensive testing of different input parameters confirmed the robustness of the result. Therefore, the thermal inertia was constrained to low values with high probability.

We propose three possible interpretations for the extremely low thermal inertia of 2016~GE1: either a high micro-porosity, or the presence of a layer of fine regolith on the surface, or the cracked surface material. Therefore this work, together with the work of \citet{2021A&A...647A..61F}, not only demonstrates the usefulness of the alternative method for constraining the thermal properties of asteroid surfaces, but also opens up the possibility of the existence of a potentially new class of NEAs with super-fast rotation and low thermal inertia. Future characterisations and in-situ explorations are needed to better understand the physical properties of such objects. In this context, the extended Hayabusa~2 mission will visit the small super-fast rotating asteroid 1998~KY26 \citep{2021AdSpR..68.1533H} and is expected to provide new insights into these very small asteroids.


\begin{acknowledgements}
   We appreciate the support from the Planetary Society STEP Grant, made possible
   by the generosity of The Planetary Society' members. 
   M. F. and B. N. also acknowledge the MSCA ETN Stardust-R, Grant Agreement n. 813644 under the
   European Union H2020 research and innovation program.
\end{acknowledgements}

\bibliographystyle{aa}
\bibliography{stardustBib.bib}{} 

\begin{thebibliography}{69}
\expandafter\ifx\csname natexlab\endcsname\relax\def\natexlab#1{#1}\fi

\bibitem[{{Al{\'\i}-Lagoa} {et~al.}(2020){Al{\'\i}-Lagoa}, {M{\"u}ller},
  {Kiss}, {Szak{\'a}ts}, {Marton}, {Farkas-Tak{\'a}cs}, {Bartczak},
  {Butkiewicz-B{\k{a}}k}, {Dudzi{\'n}ski}, {Marciniak}, {Podlewska-Gaca},
  {Duffard}, {Santos-Sanz}, \& {Ortiz}}]{2020A&A...638A..84A}
{Al{\'\i}-Lagoa}, V., {M{\"u}ller}, T.~G., {Kiss}, C., {et~al.} 2020, \aap,
  638, A84

\bibitem[{{Ballouz} {et~al.}(2020){Ballouz}, {Walsh}, {Barnouin},
  {DellaGiustina}, {Asad}, {Jawin}, {Daly}, {Bottke}, {Michel}, {Avdellidou},
  {Delbo}, {Daly}, {Asphaug}, {Bennett}, {Bierhaus}, {Connolly}, {Golish},
  {Molaro}, {Nolan}, {Pajola}, {Rizk}, {Schwartz}, {Trang}, {Wolner}, \&
  {Lauretta}}]{2020Natur.587..205B}
{Ballouz}, R.~L., {Walsh}, K.~J., {Barnouin}, O.~S., {et~al.} 2020, \nat, 587,
  205

\bibitem[{{Berthier} {et~al.}(2023){Berthier}, {Carry}, {Mahlke}, \&
  {Normand}}]{2023A&A...671A.151B}
{Berthier}, J., {Carry}, B., {Mahlke}, M., \& {Normand}, J. 2023, \aap, 671,
  A151

\bibitem[{{Bottke} {et~al.}(2006){Bottke}, {Vokrouhlick{\'y}}, {Rubincam}, \&
  {Nesvorn{\'y}}}]{bottke-etal_2006}
{Bottke}, William~F., J., {Vokrouhlick{\'y}}, D., {Rubincam}, D.~P., \&
  {Nesvorn{\'y}}, D. 2006, Annual Review of Earth and Planetary Sciences, 34,
  157

\bibitem[{{Bowell} {et~al.}(1989){Bowell}, {Hapke}, {Domingue}, {Lumme},
  {Peltoniemi}, \& {Harris}}]{bowell-etal_1989}
{Bowell}, E., {Hapke}, B., {Domingue}, D., {et~al.} 1989, in Asteroids II, ed.
  R.~P. {Binzel}, T.~{Gehrels}, \& M.~S. {Matthews}, 524--556

\bibitem[{{Cambioni} {et~al.}(2021){Cambioni}, {Delbo}, {Poggiali},
  {Avdellidou}, {Ryan}, {Deshapriya}, {Asphaug}, {Ballouz}, {Barucci},
  {Bennett}, {Bottke}, {Brucato}, {Burke}, {Cloutis}, {DellaGiustina}, {Emery},
  {Rozitis}, {Walsh}, \& {Lauretta}}]{2021Natur.598...49C}
{Cambioni}, S., {Delbo}, M., {Poggiali}, G., {et~al.} 2021, \nat, 598, 49

\bibitem[{{Carpino} {et~al.}(2003){Carpino}, {Milani}, \&
  {Chesley}}]{2003Icar..166..248C}
{Carpino}, M., {Milani}, A., \& {Chesley}, S.~R. 2003, \icarus, 166, 248

\bibitem[{{Daly} {et~al.}(2020){Daly}, {Barnouin}, {Seabrook}, {Roberts},
  {Dickinson}, {Walsh}, {Jawin}, {Palmer}, {Gaskell}, {Weirich}, {Haltigin},
  {Gaudreau}, {Brunet}, {Cunningham}, {Michel}, {Zhang}, {Ballouz}, {Neumann},
  {Perry}, {Philpott}, {Al Asad}, {Johnson}, {Adam}, {Leonard}, {Geeraert},
  {Getzandanner}, {Nolan}, {Daly}, {Bierhaus}, {Mazarico}, {Rozitis}, {Ryan},
  {Dellaguistina}, {Rizk}, {Susorney}, {Enos}, \&
  {Lauretta}}]{2020SciA....6.3649D}
{Daly}, M.~G., {Barnouin}, O.~S., {Seabrook}, J.~A., {et~al.} 2020, Science
  Advances, 6, eabd3649

\bibitem[{{Del Vigna} {et~al.}(2018){Del Vigna}, {Faggioli}, {Milani}, {Spoto},
  {Farnocchia}, \& {Carry}}]{2018A&A...617A..61D}
{Del Vigna}, A., {Faggioli}, L., {Milani}, A., {et~al.} 2018, \aap, 617, A61

\bibitem[{{Delbo'} {et~al.}(2007){Delbo'}, {dell'Oro}, {Harris}, {Mottola}, \&
  {Mueller}}]{delbo-etal_2007}
{Delbo'}, M., {dell'Oro}, A., {Harris}, A.~W., {Mottola}, S., \& {Mueller}, M.
  2007, Icarus, 190, 236

\bibitem[{{Delbo'} {et~al.}(2015){Delbo'}, {Mueller}, {Emery}, {Rozitis}, \&
  {Capria}}]{delbo-etal_2015}
{Delbo'}, M., {Mueller}, M., {Emery}, J.~P., {Rozitis}, B., \& {Capria}, M.~T.
  2015, {Asteroid Thermophysical Modeling} (University of Arizona Press),
  107--128

\bibitem[{{Delbo} {et~al.}(2022){Delbo}, {Walsh}, {Matonti}, {Wilkerson},
  {Pajola}, {Al Asad}, {Avdellidou}, {Ballouz}, {Bennett}, {Connolly},
  {DellaGiustina}, {Golish}, {Molaro}, {Rizk}, {Schwartz}, \&
  {Lauretta}}]{2022NatGe..15..453D}
{Delbo}, M., {Walsh}, K.~J., {Matonti}, C., {et~al.} 2022, Nature Geoscience,
  15, 453

\bibitem[{{Dellagiustina} {et~al.}(2019){Dellagiustina}, {Emery}, {Golish},
  {Rozitis}, {Bennett}, {Burke}, {Ballouz}, {Becker}, {Christensen}, {Drouet
  D'Aubigny}, {Hamilton}, {Reuter}, {Rizk}, {Simon}, {Asphaug}, {Bandfield},
  {Barnouin}, {Barucci}, {Bierhaus}, {Binzel}, {Bottke}, {Bowles}, {Campins},
  {Clark}, {Clark}, {Connolly}, {Daly}, {Leon}, {Delbo'}, {Deshapriya},
  {Elder}, {Fornasier}, {Hergenrother}, {Howell}, {Jawin}, {Kaplan}, {Kareta},
  {Le Corre}, {Li}, {Licandro}, {Lim}, {Michel}, {Molaro}, {Nolan}, {Pajola},
  {Popescu}, {Garcia}, {Ryan}, {Schwartz}, {Shultz}, {Siegler}, {Smith},
  {Tatsumi}, {Thomas}, {Walsh}, {Wolner}, {Zou}, {Lauretta}, \& {Osiris-Rex
  Team}}]{2019NatAs...3..341D}
{Dellagiustina}, D.~N., {Emery}, J.~P., {Golish}, D.~R., {et~al.} 2019, Nature
  Astronomy, 3, 341

\bibitem[{{Emery} {et~al.}(2014){Emery}, {Fern{\'a}ndez}, {Kelley}, {Warden},
  {Hergenrother}, {Lauretta}, {Drake}, {Campins}, \&
  {Ziffer}}]{2014Icar..234...17E}
{Emery}, J.~P., {Fern{\'a}ndez}, Y.~R., {Kelley}, M.~S.~P., {et~al.} 2014,
  \icarus, 234, 17

\bibitem[{{Farinella} {et~al.}(1998){Farinella}, {Vokrouhlick{\'y}}, \&
  {Hartmann}}]{farinella-etal_1998}
{Farinella}, P., {Vokrouhlick{\'y}}, D., \& {Hartmann}, W.~K. 1998, Icarus,
  132, 378

\bibitem[{{Farnocchia} {et~al.}(2021){Farnocchia}, {Chesley}, {Takahashi},
  {Rozitis}, {Vokrouhlick{\'y}}, {Rush}, {Mastrodemos}, {Kennedy}, {Park},
  {Bellerose}, {Lubey}, {Velez}, {Davis}, {Emery}, {Leonard}, {Geeraert},
  {Antreasian}, \& {Lauretta}}]{2021Icar..36914594F}
{Farnocchia}, D., {Chesley}, S.~R., {Takahashi}, Y., {et~al.} 2021, \icarus,
  369, 114594

\bibitem[{{Farnocchia} {et~al.}(2013){Farnocchia}, {Chesley},
  {Vokrouhlick{\'y}}, {Milani}, {Spoto}, \& {Bottke}}]{farnocchia-etal_2013}
{Farnocchia}, D., {Chesley}, S.~R., {Vokrouhlick{\'y}}, D., {et~al.} 2013,
  \icarus, 224, 1

\bibitem[{{Farnocchia} {et~al.}(2023){Farnocchia}, {Seligman}, {Granvik},
  {Hainaut}, {Meech}, {Micheli}, {Weryk}, {Chesley}, {Christensen}, {Koschny},
  {Kleyna}, {Lazzaro}, {Mommert}, \& {Wainscoat}}]{2023PSJ.....4...29F}
{Farnocchia}, D., {Seligman}, D.~Z., {Granvik}, M., {et~al.} 2023, \psj, 4, 29

\bibitem[{{Fenucci} {et~al.}(2021){Fenucci}, {Novakovi{\'c}},
  {Vokrouhlick{\'y}}, \& {Weryk}}]{2021A&A...647A..61F}
{Fenucci}, M., {Novakovi{\'c}}, B., {Vokrouhlick{\'y}}, D., \& {Weryk}, R.~J.
  2021, \aap, 647, A61

\bibitem[{{Flynn} {et~al.}(2018){Flynn}, {Consolmagno}, {Brown}, \&
  {Macke}}]{2018ChEG...78..269F}
{Flynn}, G.~J., {Consolmagno}, G.~J., {Brown}, P., \& {Macke}, R.~J. 2018,
  Chemie der Erde / Geochemistry, 78, 269

\bibitem[{{Folkner} {et~al.}(2014){Folkner}, {Williams}, {Boggs}, {Park}, \&
  {Kuchynka}}]{2014IPNPR.196C...1F}
{Folkner}, W.~M., {Williams}, J.~G., {Boggs}, D.~H., {Park}, R.~S., \&
  {Kuchynka}, P. 2014, Interplanetary Network Progress Report, 42-196, 1

\bibitem[{{Ghosal} {et~al.}(2022){Ghosal}, {Jedicke}, \&
  {Bolin}}]{2022DPS....5452306G}
{Ghosal}, M., {Jedicke}, R., \& {Bolin}, B. 2022, in AAS/Division for Planetary
  Sciences Meeting Abstracts, Vol.~54, AAS/Division for Planetary Sciences
  Meeting Abstracts, 523.06

\bibitem[{Granvik {et~al.}(2018)Granvik, Morbidelli, Jedicke, Bolin, Bottke,
  Beshore, Vokrouhlický, Nesvorný, \& Michel}]{granvik-etal_2018}
Granvik, M., Morbidelli, A., Jedicke, R., {et~al.} 2018, Icarus, 312, 181

\bibitem[{{Greenberg} {et~al.}(2020){Greenberg}, {Margot}, {Verma}, {Taylor},
  \& {Hodge}}]{greenberg-etal_2020}
{Greenberg}, A.~H., {Margot}, J.-L., {Verma}, A.~K., {Taylor}, P.~A., \&
  {Hodge}, S.~E. 2020, The astronomical Journal, 159, 92

\bibitem[{{Grott} {et~al.}(2019){Grott}, {Knollenberg}, {Hamm}, {Ogawa},
  {Jaumann}, {Otto}, {Delbo}, {Michel}, {Biele}, {Neumann}, {Knapmeyer},
  {K{\"u}hrt}, {Senshu}, {Okada}, {Helbert}, {Maturilli}, {M{\"u}ller},
  {Hagermann}, {Sakatani}, {Tanaka}, {Arai}, {Mottola}, {Tachibana}, {Pelivan},
  {Drube}, {Vincent}, {Yano}, {Pilorget}, {Matz}, {Schmitz}, {Koncz},
  {Schr{\"o}der}, {Trauthan}, {Schlotterer}, {Krause}, {Ho}, \&
  {Moussi-Soffys}}]{grott-etal_2020}
{Grott}, M., {Knollenberg}, J., {Hamm}, M., {et~al.} 2019, Nature Astronomy, 3,
  971

\bibitem[{{Groussin} {et~al.}(2019){Groussin}, {Attree}, {Brouet}, {Ciarletti},
  {Davidsson}, {Filacchione}, {Fischer}, {Gundlach}, {Knapmeyer},
  {Knollenberg}, {Kokotanekova}, {K{\"u}hrt}, {Leyrat}, {Marshall}, {Pelivan},
  {Skorov}, {Snodgrass}, {Spohn}, \& {Tosi}}]{2019SSRv..215...29G}
{Groussin}, O., {Attree}, N., {Brouet}, Y., {et~al.} 2019, \ssr, 215, 29

\bibitem[{{Hamm} {et~al.}(2022){Hamm}, {Grott}, {Senshu}, {Knollenberg}, {de
  Wiljes}, {Hamilton}, {Scholten}, {Matz}, {Bates}, {Maturilli}, {Shimaki},
  {Sakatani}, {Neumann}, {Okada}, {Preusker}, {Elgner}, {Helbert}, {K{\"u}hrt},
  {Ho}, {Tanaka}, {Jaumann}, \& {Sugita}}]{2022NatCo..13..364H}
{Hamm}, M., {Grott}, M., {Senshu}, H., {et~al.} 2022, Nature Communications,
  13, 364

\bibitem[{{Harris} \& {Drube}(2016)}]{harris-drube_2016}
{Harris}, A.~W. \& {Drube}, L. 2016, The Astrophysical Journal, 832, 127

\bibitem[{{Henke} {et~al.}(2016){Henke}, {Gail}, \&
  {Trieloff}}]{2016A&A...589A..41H}
{Henke}, S., {Gail}, H.-P., \& {Trieloff}, M. 2016, \aap, 589, A41

\bibitem[{{Hergenrother} {et~al.}(2019){Hergenrother}, {Maleszewski}, {Nolan},
  {Li}, {Drouet D'Aubigny}, {Shelly}, {Howell}, {Kareta}, {Izawa}, {Barucci},
  {Bierhaus}, {Campins}, {Chesley}, {Clark}, {Christensen}, {Dellagiustina},
  {Fornasier}, {Golish}, {Hartzell}, {Rizk}, {Scheeres}, {Smith}, {Zou},
  {Lauretta}, \& {OSIRIS-REx Team}}]{2019NatCo..10.1291H}
{Hergenrother}, C.~W., {Maleszewski}, C.~K., {Nolan}, M.~C., {et~al.} 2019,
  Nature Communications, 10, 1291

\bibitem[{{Hirabayashi} {et~al.}(2021){Hirabayashi}, {Mimasu}, {Sakatani},
  {Watanabe}, {Tsuda}, {Saiki}, {Kikuchi}, {Kouyama}, {Yoshikawa}, {Tanaka},
  {Nakazawa}, {Takei}, {Terui}, {Takeuchi}, {Fujii}, {Iwata}, {Tsumura},
  {Matsuura}, {Shimaki}, {Urakawa}, {Ishibashi}, {Hasegawa}, {Ishiguro},
  {Kuroda}, {Okumura}, {Sugita}, {Okada}, {Kameda}, {Kamata}, {Higuchi},
  {Senshu}, {Noda}, {Matsumoto}, {Suetsugu}, {Hirai}, {Kitazato}, {Farnocchia},
  {Naidu}, {Tholen}, {Hergenrother}, {Whiteley}, {Moskovitz}, {Abell}, \& {the
  Hayabusa2 extended mission study Group}}]{2021AdSpR..68.1533H}
{Hirabayashi}, M., {Mimasu}, Y., {Sakatani}, N., {et~al.} 2021, Advances in
  Space Research, 68, 1533

\bibitem[{{Ho} {et~al.}(2017){Ho}, {Baturkin}, {Grimm}, {Grundmann}, {Hobbie},
  {Ksenik}, {Lange}, {Sasaki}, {Schlotterer}, {Talapina}, {Termtanasombat},
  {Wejmo}, {Witte}, {Wrasmann}, {W{\"u}bbels}, {R{\"o}{\ss}ler}, {Ziach},
  {Findlay}, {Biele}, {Krause}, {Ulamec}, {Lange}, {Mierheim}, {Lichtenheldt},
  {Maier}, {Reill}, {Sedlmayr}, {Bousquet}, {Bellion}, {Bompis},
  {Cenac-Morthe}, {Deleuze}, {Fredon}, {Jurado}, {Canalias}, {Jaumann},
  {Bibring}, {Glassmeier}, {Hercik}, {Grott}, {Celotti}, {Cordero},
  {Hendrikse}, \& {Okada}}]{2017SSRv..208..339H}
{Ho}, T.-M., {Baturkin}, V., {Grimm}, C., {et~al.} 2017, \ssr, 208, 339

\bibitem[{{Hu} {et~al.}(2021){Hu}, {Richardson}, {Zhang}, \&
  {Ji}}]{2021MNRAS.502.5277H}
{Hu}, S., {Richardson}, D.~C., {Zhang}, Y., \& {Ji}, J. 2021, \mnras, 502, 5277

\bibitem[{{Hung} {et~al.}(2022){Hung}, {Hanu{\v{s}}}, {Masiero}, \&
  {Tholen}}]{2022PSJ.....3...56H}
{Hung}, D., {Hanu{\v{s}}}, J., {Masiero}, J.~R., \& {Tholen}, D.~J. 2022, \psj,
  3, 56

\bibitem[{{Ishizaki} {et~al.}(2023){Ishizaki}, {Nagano}, {Tanaka}, {Sakatani},
  {Nakamura}, {Okada}, {Fujita}, {Alasli}, {Morita}, {Kikuiri}, {Amano},
  {Kagawa}, {Yurimoto}, {Noguchi}, {Okazaki}, {Yabuta}, {Naraoka}, {Sakamoto},
  {Tachibana}, {Watanabe}, \& {Tsuda}}]{2023IJT....44...51I}
{Ishizaki}, T., {Nagano}, H., {Tanaka}, S., {et~al.} 2023, International
  Journal of Thermophysics, 44, 51

\bibitem[{{Lauretta} {et~al.}(2019){Lauretta}, {Dellagiustina}, {Bennett},
  {Golish}, {Becker}, {Balram-Knutson}, {Barnouin}, {Becker}, {Bottke},
  {Boynton}, {Campins}, {Clark}, {Connolly}, {Drouet D'Aubigny}, {Dworkin},
  {Emery}, {Enos}, {Hamilton}, {Hergenrother}, {Howell}, {Izawa}, {Kaplan},
  {Nolan}, {Rizk}, {Roper}, {Scheeres}, {Smith}, {Walsh}, {Wolner}, \&
  {Osiris-Rex Team}}]{lauretta-etal_2019}
{Lauretta}, D.~S., {Dellagiustina}, D.~N., {Bennett}, C.~A., {et~al.} 2019,
  \nat, 568, 55

\bibitem[{{Leyrat} {et~al.}(2012){Leyrat}, {Barucci}, {Mueller}, {O'Rourke},
  {Valtchanov}, \& {Fornasier}}]{2012A&A...539A.154L}
{Leyrat}, C., {Barucci}, A., {Mueller}, T., {et~al.} 2012, \aap, 539, A154

\bibitem[{{MacLennan} \& {Emery}(2021)}]{2021PSJ.....2..161M}
{MacLennan}, E.~M. \& {Emery}, J.~P. 2021, \psj, 2, 161

\bibitem[{{Marciniak} {et~al.}(2019){Marciniak}, {Al{\'\i}-Lagoa},
  {M{\"u}ller}, {Szak{\'a}ts}, {Moln{\'a}r}, {P{\'a}l}, {Podlewska-Gaca},
  {Parley}, {Antonini}, {Barbotin}, {Behrend}, {Bernasconi},
  {Butkiewicz-B{\k{a}}k}, {Crippa}, {Duffard}, {Ditteon}, {Feuerbach},
  {Fauvaud}, {Garlitz}, {Geier}, {Goncalves}, {Grice}, {Grze{\'s}kowiak},
  {Hirsch}, {Horbowicz}, {Kami{\'n}ski}, {Kami{\'n}ska}, {Kim}, {Kim},
  {Konstanciak}, {Kudak}, {Kulczak}, {Maestre}, {Manzini}, {Marks}, {Monteiro},
  {Og{\l}oza}, {Oszkiewicz}, {Pilcher}, {Perig}, {Polakis}, {Poli{\'n}ska},
  {Roy}, {Sanabria}, {Santana-Ros}, {Skiff}, {Skrzypek}, {Sobkowiak}, {Sonbas},
  {Thizy}, {Trela}, {Urakawa}, {{\.Z}ejmo}, \&
  {{\.Z}ukowski}}]{2019A&A...625A.139M}
{Marciniak}, A., {Al{\'\i}-Lagoa}, V., {M{\"u}ller}, T.~G., {et~al.} 2019,
  \aap, 625, A139

\bibitem[{{Marshall} {et~al.}(2018){Marshall}, {Groussin}, {Vincent}, {Brouet},
  {Kappel}, {Arnold}, {Capria}, {Filacchione}, {Hartogh}, {Hofstadter}, {Ip},
  {Jorda}, {K{\"u}hrt}, {Lellouch}, {Mottola}, {Rezac}, {Rodrigo}, {Rodionov},
  {Schloerb}, \& {Thomas}}]{2018A&A...616A.122M}
{Marshall}, D., {Groussin}, O., {Vincent}, J.~B., {et~al.} 2018, \aap, 616,
  A122

\bibitem[{Milani \& Gronchi(2009)}]{milani-gronchi_2009}
Milani, A. \& Gronchi, G.~F. 2009, Theory of Orbit Determination (Cambridge
  University Press)

\bibitem[{{Molaro} {et~al.}(2020){Molaro}, {Walsh}, {Jawin}, {Ballouz},
  {Bennett}, {DellaGiustina}, {Golish}, {Drouet d'Aubigny}, {Rizk}, {Schwartz},
  {Hanna}, {Martel}, {Pajola}, {Campins}, {Ryan}, {Bottke}, \&
  {Lauretta}}]{molaro-etal_2020}
{Molaro}, J.~L., {Walsh}, K.~J., {Jawin}, E.~R., {et~al.} 2020, Nature
  Communications, 11, 2913

\bibitem[{{Morbidelli} {et~al.}(2020){Morbidelli}, {Delbo}, {Granvik},
  {Bottke}, {Jedicke}, {Bolin}, {Michel}, \&
  {Vokrouhlicky}}]{morbidelli-etal_2020}
{Morbidelli}, A., {Delbo}, M., {Granvik}, M., {et~al.} 2020, Icarus, 340,
  113631

\bibitem[{{Murdoch} {et~al.}(2021){Murdoch}, {Drilleau}, {Sunday}, {Thuillet},
  {Wilhelm}, {Nguyen}, \& {Gourinat}}]{2021MNRAS.503.3460M}
{Murdoch}, N., {Drilleau}, M., {Sunday}, C., {et~al.} 2021, \mnras, 503, 3460

\bibitem[{{Novakovi{\'c}} {et~al.}(2022){Novakovi{\'c}}, {Vokrouhlick{\'y}},
  {Spoto}, \& {Nesvorn{\'y}}}]{2022CeMDA.134...34N}
{Novakovi{\'c}}, B., {Vokrouhlick{\'y}}, D., {Spoto}, F., \& {Nesvorn{\'y}}, D.
  2022, Celestial Mechanics and Dynamical Astronomy, 134, 34

\bibitem[{{Ostrowski} \& {Bryson}(2019)}]{ostrowsky-bryson_2019}
{Ostrowski}, D. \& {Bryson}, K. 2019, Planetary and Space Science, 165, 148

\bibitem[{{Persson} \& {Biele}(2021)}]{2021arXiv211015258P}
{Persson}, B.~N.~J. \& {Biele}, J. 2021, arXiv e-prints, arXiv:2110.15258

\bibitem[{{Petkovi{\'c}} {et~al.}(2021){Petkovi{\'c}}, {Fenucci}, \&
  {Novakovi{\'c}}}]{2021EPSC...15..390P}
{Petkovi{\'c}}, V., {Fenucci}, M., \& {Novakovi{\'c}}, B. 2021, in European
  Planetary Science Congress, EPSC2021--390

\bibitem[{{Piqueux} {et~al.}(2021){Piqueux}, {Vu}, {Bapst}, {Garvie},
  {Choukroun}, \& {Edwards}}]{2021JGRE..12607003P}
{Piqueux}, S., {Vu}, T.~H., {Bapst}, J., {et~al.} 2021, Journal of Geophysical
  Research (Planets), 126, e07003

\bibitem[{{Pravec} \& {Harris}(2000)}]{pravec-harris_2000}
{Pravec}, P. \& {Harris}, A.~W. 2000, Icarus, 148, 12

\bibitem[{{Pravec} \& {Harris}(2007)}]{2007Icar..190..250P}
{Pravec}, P. \& {Harris}, A.~W. 2007, \icarus, 190, 250

\bibitem[{{Rognini} {et~al.}(2020){Rognini}, {Capria}, {Tosi}, {De Sanctis},
  {Ciarniello}, {Longobardo}, {Carrozzo}, {Raponi}, {Frigeri}, {Palomba},
  {Fonte}, {Giardino}, {Ammannito}, {Raymond}, \&
  {Russell}}]{2020JGRE..12505733R}
{Rognini}, E., {Capria}, M.~T., {Tosi}, F., {et~al.} 2020, Journal of
  Geophysical Research (Planets), 125, e05733

\bibitem[{{Rozitis} \& {Green}(2011)}]{2011MNRAS.415.2042R}
{Rozitis}, B. \& {Green}, S.~F. 2011, \mnras, 415, 2042

\bibitem[{{Rozitis} \& {Green}(2014)}]{2014A&A...568A..43R}
{Rozitis}, B. \& {Green}, S.~F. 2014, \aap, 568, A43

\bibitem[{{Rozitis} {et~al.}(2020){Rozitis}, {Ryan}, {Emery}, {Christensen},
  {Hamilton}, {Simon}, {Reuter}, {Al Asad}, {Ballouz}, {Bandfield}, {Barnouin},
  {Bennett}, {Bernacki}, {Burke}, {Cambioni}, {Clark}, {Daly}, {Delbo},
  {DellaGiustina}, {Elder}, {Hanna}, {Haberle}, {Howell}, {Golish}, {Jawin},
  {Kaplan}, {Lim}, {Molaro}, {Munoz}, {Nolan}, {Rizk}, {Siegler}, {Susorney},
  {Walsh}, \& {Lauretta}}]{2020SciA....6.3699R}
{Rozitis}, B., {Ryan}, A.~J., {Emery}, J.~P., {et~al.} 2020, Science Advances,
  6, eabc3699

\bibitem[{{Sakatani} {et~al.}(2021){Sakatani}, {Tanaka}, {Okada}, {Fukuhara},
  {Riu}, {Sugita}, {Honda}, {Morota}, {Kameda}, {Yokota}, {Tatsumi}, {Yumoto},
  {Hirata}, {Miura}, {Kouyama}, {Senshu}, {Shimaki}, {Arai}, {Takita},
  {Demura}, {Sekiguchi}, {M{\"u}ller}, {Hagermann}, {Biele}, {Grott}, {Hamm},
  {Delbo}, {Neumann}, {Taguchi}, {Ogawa}, {Matsunaga}, {Wada}, {Hasegawa},
  {Helbert}, {Hirata}, {Noguchi}, {Yamada}, {Suzuki}, {Honda}, {Ogawa},
  {Hayakawa}, {Yoshioka}, {Matsuoka}, {Cho}, {Sawada}, {Kitazato}, {Iwata},
  {Abe}, {Ohtake}, {Matsuura}, {Matsumoto}, {Noda}, {Ishihara}, {Yamamoto},
  {Higuchi}, {Namiki}, {Ono}, {Saiki}, {Imamura}, {Takagi}, {Yano}, {Shirai},
  {Okamoto}, {Nakazawa}, {Iijima}, {Arakawa}, {Wada}, {Kadono}, {Ishibashi},
  {Terui}, {Kikuchi}, {Yamaguchi}, {Ogawa}, {Mimasu}, {Yoshikawa}, {Takahashi},
  {Takei}, {Fujii}, {Takeuchi}, {Yamamoto}, {Hirose}, {Hosoda}, {Mori},
  {Shimada}, {Soldini}, {Tsukizaki}, {Ozaki}, {Tachibana}, {Ikeda}, {Ishiguro},
  {Yabuta}, {Yoshikawa}, {Watanabe}, \& {Tsuda}}]{2021NatAs...5..766S}
{Sakatani}, N., {Tanaka}, S., {Okada}, T., {et~al.} 2021, Nature Astronomy, 5,
  766

\bibitem[{{S{\'a}nchez} \& {Scheeres}(2020)}]{sanchez-scheeres_2020}
{S{\'a}nchez}, P. \& {Scheeres}, D.~J. 2020, \icarus, 338, 113443

\bibitem[{{Shimaki} {et~al.}(2020){Shimaki}, {Senshu}, {Sakatani}, {Okada},
  {Fukuhara}, {Tanaka}, {Taguchi}, {Arai}, {Demura}, {Ogawa}, {Suko},
  {Sekiguchi}, {Kouyama}, {Hasegawa}, {Takita}, {Matsunaga}, {Imamura}, {Wada},
  {Kitazato}, {Hirata}, {Hirata}, {Noguchi}, {Sugita}, {Kikuchi}, {Yamaguchi},
  {Ogawa}, {Ono}, {Mimasu}, {Yoshikawa}, {Takahashi}, {Takei}, {Fujii},
  {Takeuchi}, {Yamamoto}, {Yamada}, {Shirai}, {Iijima}, {Ogawa}, {Nakazawa},
  {Terui}, {Saiki}, {Yoshikawa}, {Tsuda}, \& {Watanabe}}]{shimaki-etal_2020}
{Shimaki}, Y., {Senshu}, H., {Sakatani}, N., {et~al.} 2020, Icarus, 348, 113835

\bibitem[{Stoer \& Bulirsch(2002)}]{bulirsch:02}
Stoer, J. \& Bulirsch, R. 2002, Introduction to numerical analysis, Texts in
  applied mathematics (Springer)

\bibitem[{{Tardioli} {et~al.}(2017){Tardioli}, {Farnocchia}, {Rozitis},
  {Cotto-Figueroa}, {Chesley}, {Statler}, \& {Vasile}}]{2017A&A...608A..61T}
{Tardioli}, C., {Farnocchia}, D., {Rozitis}, B., {et~al.} 2017, \aap, 608, A61

\bibitem[{{Thomas} {et~al.}(2011){Thomas}, {Trilling}, {Emery}, {Mueller},
  {Hora}, {Benner}, {Bhattacharya}, {Bottke}, {Chesley}, {Delb{\'o}}, {Fazio},
  {Harris}, {Mainzer}, {Mommert}, {Morbidelli}, {Penprase}, {Smith}, {Spahr},
  \& {Stansberry}}]{2011AJ....142...85T}
{Thomas}, C.~A., {Trilling}, D.~E., {Emery}, J.~P., {et~al.} 2011, \aj, 142, 85

\bibitem[{{Usui} {et~al.}(2013){Usui}, {Kasuga}, {Hasegawa}, {Ishiguro},
  {Kuroda}, {M{\"u}ller}, {Ootsubo}, \& {Matsuhara}}]{2013ApJ...762...56U}
{Usui}, F., {Kasuga}, T., {Hasegawa}, S., {et~al.} 2013, \apj, 762, 56

\bibitem[{{Vokrouhlick{\'y}}(1998)}]{vokrouhlicky_1998}
{Vokrouhlick{\'y}}, D. 1998, \aap, 338, 353

\bibitem[{{Vokrouhlick{\'y}}(1999)}]{1999A&A...344..362V}
{Vokrouhlick{\'y}}, D. 1999, \aap, 344, 362

\bibitem[{{Vokrouhlick{\'y}} {et~al.}(2017){Vokrouhlick{\'y}}, {Pravec},
  {{\v{D}}urech}, {Hornoch}, {Ku{\v{s}}nir{\'a}k}, {Gal{\'a}d},
  {Vra{\v{s}}til}, {Ku{\v{c}}{\'a}kov{\'a}}, {Pollock}, {Ortiz}, {Morales},
  {Gaftonyuk}, {Pray}, {Krugly}, {Inasaridze}, {Ayvazian}, {Molotov}, \&
  {Colazo}}]{vokrouhlicky-etal_2017}
{Vokrouhlick{\'y}}, D., {Pravec}, P., {{\v{D}}urech}, J., {et~al.} 2017, \aj,
  153, 270

\bibitem[{{Warner}(2016)}]{2016MPBu...43..240W}
{Warner}, B.~D. 2016, Minor Planet Bulletin, 43, 240

\bibitem[{{Warner} {et~al.}(2009){Warner}, {Harris}, \&
  {Pravec}}]{warner-etal_2009}
{Warner}, B.~D., {Harris}, A.~W., \& {Pravec}, P. 2009, Icarus, 202, 134

\bibitem[{{Watanabe} {et~al.}(2019){Watanabe}, {Hirabayashi}, {Hirata},
  {Hirata}, {Noguchi}, {Shimaki}, {Ikeda}, {Tatsumi}, {Yoshikawa}, {Kikuchi},
  {Yabuta}, {Nakamura}, {Tachibana}, {Ishihara}, {Morota}, {Kitazato},
  {Sakatani}, {Matsumoto}, {Wada}, {Senshu}, {Honda}, {Michikami}, {Takeuchi},
  {Kouyama}, {Honda}, {Kameda}, {Fuse}, {Miyamoto}, {Komatsu}, {Sugita},
  {Okada}, {Namiki}, {Arakawa}, {Ishiguro}, {Abe}, {Gaskell}, {Palmer},
  {Barnouin}, {Michel}, {French}, {McMahon}, {Scheeres}, {Abell}, {Yamamoto},
  {Tanaka}, {Shirai}, {Matsuoka}, {Yamada}, {Yokota}, {Suzuki}, {Yoshioka},
  {Cho}, {Tanaka}, {Nishikawa}, {Sugiyama}, {Kikuchi}, {Hemmi}, {Yamaguchi},
  {Ogawa}, {Ono}, {Mimasu}, {Yoshikawa}, {Takahashi}, {Takei}, {Fujii},
  {Hirose}, {Iwata}, {Hayakawa}, {Hosoda}, {Mori}, {Sawada}, {Shimada},
  {Soldini}, {Yano}, {Tsukizaki}, {Ozaki}, {Iijima}, {Ogawa}, {Fujimoto}, {Ho},
  {Moussi}, {Jaumann}, {Bibring}, {Krause}, {Terui}, {Saiki}, {Nakazawa}, \&
  {Tsuda}}]{watanabe-etal_2019}
{Watanabe}, S., {Hirabayashi}, M., {Hirata}, N., {et~al.} 2019, Science, 364,
  268

\bibitem[{{Zhang} {et~al.}(2021){Zhang}, {Michel}, {Richardson}, {Barnouin},
  {Agrusa}, {Tsiganis}, {Manzoni}, \& {May}}]{2021Icar..36214433Z}
{Zhang}, Y., {Michel}, P., {Richardson}, D.~C., {et~al.} 2021, \icarus, 362,
  114433

\end{thebibliography}
\end{document}